\documentclass[twocolumn,amsmath,amssymb,superscriptaddress,pra,a4paper,floatfix]{revtex4}
\usepackage{graphicx}
\pdfoutput=1 
\usepackage[paperwidth=212mm,paperheight=297mm,centering,hmargin=1.65cm,vmargin=2.6cm]{geometry}
\usepackage{graphicx}
\usepackage{amsmath}
\usepackage{amssymb}
\usepackage{comment}
\usepackage{bm}
\usepackage{bbm}
\usepackage{verbatim}
\usepackage{physics}
\graphicspath{{graphics/}}
\def\r1{\textbf{r}}

\newcommand{\ba}{\begin{eqnarray}}
\newcommand{\ea}{\end{eqnarray}}
\newcommand{\ban}{\begin{eqnarray*}}
\newcommand{\ean}{\end{eqnarray*}}
\newcommand{\be}{\begin{equation}}
\newcommand{\ee}{\end{equation}}
\newcommand{\bracket}[3]{\langle#1|#2|#3\rangle}
\newcommand{\expect}[1]{\langle#1\rangle}

\begin{document}

\title{Coherent scattering from coupled two level systems}
\author{Thomas Nutz}

\affiliation{Controlled Quantum Dynamics Theory Group, Imperial College London, London SW7 2AZ, United Kingdom}
\affiliation{Quantum Engineering Technology Labs, H. H. Wills Physics Laboratory and Department of Electrical and Electronic Engineering, 
University of Bristol, BS8 1FD, UK}

\author{Samuel T. Mister}
\email{samuel.mister@bristol.ac.uk}
\altaffiliation{These authors contributed equally to this work}

\affiliation{Quantum Engineering Technology Labs and Quantum Engineering Centre for Doctoral Training, H. H. Wills Physics Laboratory and Department of Electrical and Electronic Engineering, 
University of Bristol, BS8 1FD, UK}

\author{Petros Androvitsaneas}
\author{Andrew Young}
\author{E. Harbord}
\author{J. G. Rarity}
\author{Ruth Oulton}
\author{Dara P. S. McCutcheon}
\affiliation{Quantum Engineering Technology Labs, H. H. Wills Physics Laboratory and Department of Electrical and Electronic Engineering, 
University of Bristol, BS8 1FD, UK}

\date{\today}

\begin{abstract}
We study the resonance fluorescence properties of an optically active spin $1/2$ system, elucidating the effects of a magnetic field on the coherence of the scattered light.
 We derive a master equation model for this system that reproduces the results of a two level system (TLS) while also being applicable to a spin system with ground state coupling. This model is then solved analytically in the weak excitation regime.
The inclusion of spin dynamics in our model alters the properties of the coherently scattered light at a fundamental level. For a TLS the coherence properties are known to be determined by the input laser. We show that spin scattered light inherits the coherence properties of the spin. This mapping allows us to measure spin dynamics and coherence time through direct measurement of the scattered fields. Furthermore, we show the ability to resolve sub-natural linewidth zeeman splittings. Along with representing an invaluable tool for spin spectroscopy understanding the coherence properties of the spin-scattered field will be vital for spin-photon based quantum technologies. 
\end{abstract}

\maketitle

\section{Introduction}
Spin--photon interactions give rise to effects and applications beyond those possible with a simple two level system, as output photons become entangled with the internal spin states. This has a myriad of applications involving entanglement distributon \cite{Humphreys-2018}, and the production of multi-photon entangled states \cite{lindner_proposal_2009,Gershoni-2023}. In particular, semiconductor quantum dot (QDs) and defect centre single photon sources have been 
shown theoretically~\cite{lindner_proposal_2009, economou_optically_2010, lee_quantum_2019, denning_protocol_2017-1} and 
now experimentally~\cite{schwartz_deterministic_2016, vasconcelos_scalable_2019, scerri_frequency-encoded_2018, delteil_generation_2016, stockill_phase-tuned_2017} to emit entangled photons. 
While some methods of conventional quantum optics have been successfully applied to these solid-state systems, 
new approaches are needed to realize the potential of this hybrid domain of quantum physics.

A crucial step in the theoretical description of these entanglement schemes is the assumption that the four-level system formed by the two states of a spin $1/2$ particle and their corresponding optically accessible excited states 
can be treated as two independent two-level schemes (TLS), as far as the light--matter interaction is concerned. 
This assumption makes it possible to use central techniques and results of quantum optics such as single-photon 
wavepacket emission~\cite{economou_unified_2005, lindner_proposal_2009, nguyen_ultra-coherent_2011}, 
the giant phase shift~\cite{hu_giant_2008, androvitsaneas_charged_2016, hofmann_optimized_2003, nutz_stabilization_2019}, or the spectrum of resonance fluorescence~\cite{mollow_power_1969, cresser1982resonance, matthiesen_subnatural_2012, proux_measuring_2015, bennett_cavity-enhanced_2016}. Such a treatment is valid in the regime of negligible coupling between the two TLS, however some amount of coupling is required in most entanglement schemes. We address this challenge by developing the theory of a simple and yet important and versatile system, namely a single spin in a magnetic field interacting with laser light. 
This theory allows us to elucidate several effects arising due to the replacement of the conventionally considered ground-excited state system with a spin system as a quantum emitter. Most importantly the presence of this spin gives rise to a new type of coherent scattering with a characteristic spectrum that we refer to as spin scattering.
The system we consider consists of two coupled two-level systems (TLS).  
This system is naturally found in charged III-V semiconductor QDs~\cite{warburton_single_2013}, where the two ground states can be identified as electron spin states and the coupling is provided by a magnetic field orthogonal to the optical axis. 
We investigate the coherence properties of the  light scattered from this system, and find it to differ qualitatively from light scattered by a single or two independent TLS emitters. Most notably, in the conventional coherent scattering regime of a single TLS, characterised 
by low excitation probability and weak laser driving, the scattered light inherits the coherence properties of the laser, 
such that a single-frequency excitation laser would lead to scattered light with a Dirac 
$\delta$-distribution-like emission spectrum~\cite{mollow_power_1969, carmichael1999statistical, GIBBS197687, volz_atom_2007}. 
This central quantum optical result, however, does not hold true for our spin system. We show the scattered light does not inherit the coherence properties of the laser, but those of the spin.
The mapping of the spin coherence on to the spin scattered field manifests as the spectrum being split into two narrow spectral lines. The separation of these lines is related to the coupling between the two ground states. Additionally the bandwidth of the individual spectral features is related to dephasing processes. We further show that even if this spin lives in a completely decoherence-free environment, the excitation laser itself acts as a source of dephasing, such that some amount of spectral broadening of the scattered light is unavoidable.

\section{Modeling the dynamics of the combined spin-photon system}
\begin{figure}
\includegraphics[width=\linewidth]{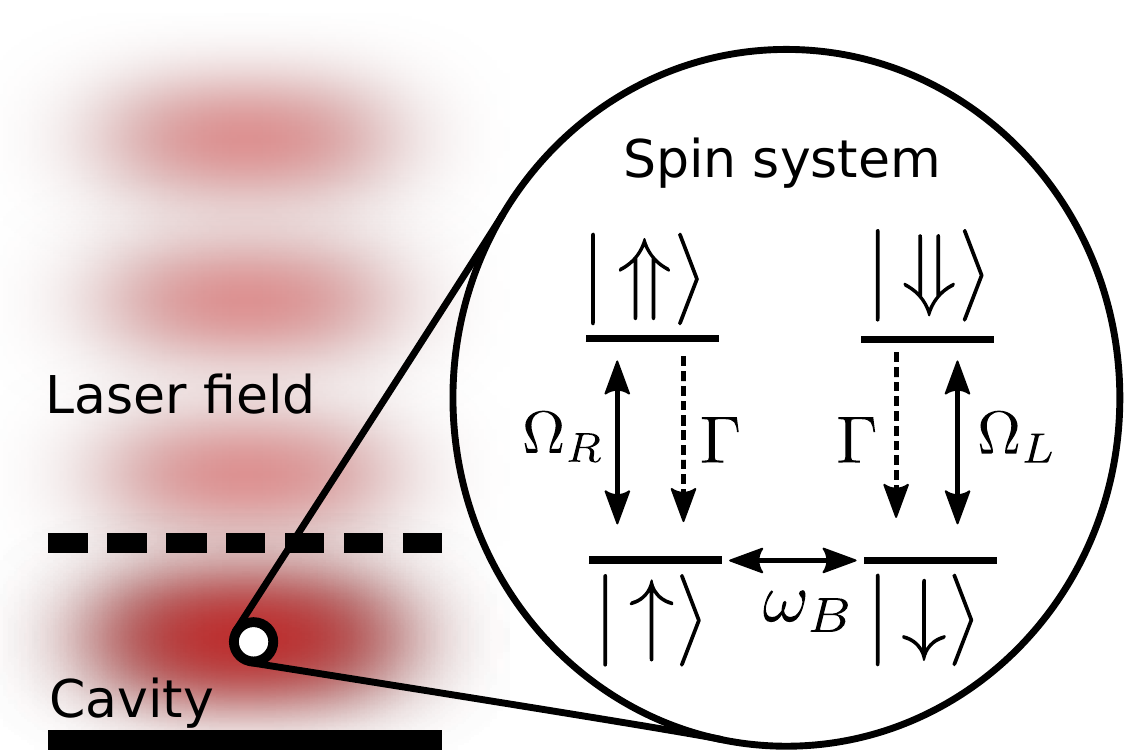}
\caption{Schematic of a spin-photon system realized by a charged quantum dot inside an optical cavity. The round inset shows a level scheme with couplings as described by the Master equation of Eq. \ref{eq: final Master equation}, where $\Omega_{R/L}$ are Rabi frequencies and $\Gamma$ denotes the Purcell-enhanced decay rate.}
 \label{fig: Faraday scheme}
\end{figure}

We frame our analysis in terms of the optically active states in a charged III-V quantum dot (QD).
The ground states constitute an effective spin $1/2$ system with basis states obeying $\sigma_z \ket{\uparrow /\downarrow} = \pm 1/2 \ket{\uparrow / \downarrow}$, where the quantisation axis $z$ with Pauli operator $\sigma_z$ is chosen along the optical and growth axis of the QD (Faraday basis). 
This spin resides in a magnetic field and is exposed to laser light via an optically driven cavity. 
The two spin ground states 
couple to two orthogonally polarised optical modes of the cavity 
which in turn couple to two orthogonally polarised continua of free-space modes. 
For concreteness we assume that the ground state $\ket{\uparrow}$ couples to a right-hand circularly polarised cavity mode 
through an excited state $\ket{\Uparrow}$, while $\ket{\downarrow}$ couples to a left-hand circularly polarised cavity mode via 
$\ket{\Downarrow}$. In a III-V QD these excited states correspond to charged excitons (trions) of angular momentum $\pm 3/2$. A schematic of this system is shown in Fig. \ref{fig: Faraday scheme}, 
and our aim is describe the effect of the laser excitation on the system, and in turn the emitted field characteristics. 

The full Hamiltonian containing spin, cavity and port mode degrees of freedom is 
\begin{equation}
    H = H^0_s + H^0_c + H^0_p + H_{cp}^I + H_{sc}^I
    \label{eq: full Hamiltonian}
\end{equation}
where (set $\hbar =1$)
\begin{equation}
\begin{split}
H^0_s &= \omega_0 P_e + \omega_B \sigma_x \\
H^0_c &= \omega_c \left( a_R^{\dagger}a_R + a_L^{\dagger} a_L \right) \\
H^0_p &= \sum_k  \omega_k \left( r_{k,R}^{\dagger} r_{k,R} + r_{k,L}^{\dagger} r_{k,L} \right)
\end{split}
 \label{eq: free Hamiltonians}
\end{equation}
with excited state energy $\omega_0$ and excited state projector $P_e = \dyad{\Uparrow}{\Uparrow} + \dyad{\Uparrow}{\Uparrow}$. The magnetic field is assumed perpendicular to the optical axis, and 
is described by the Zeeman term 
$\omega_B \sigma_x = \omega_B \left( \dyad{\uparrow}{\downarrow} + \dyad{\downarrow}{\uparrow}\right)$. 
The right- and left-handed cavity modes with annihilation operators $a_R$ and $a_L$ have photon energy $\omega_c$, 
and each couples to a quasi-continuum of port modes with annihilation operators $r_{k,R}$ and $r_{k,L}$, with energies $\omega_k$. 
The interaction part of the Hamiltonian is given by 
\begin{equation}
\begin{split}
H_{cp}^I &= \sum_k \kappa^*_k \left( r_{k,R} a_R^{\dagger} + r_{k,L} a_L^{\dagger} \right) + \mathrm{H.c.} \\
H_{sc}^I &= g \left(S_-^R a_R^{\dagger} + S_-^L a_L^{\dagger} + \mathrm{H.c.} \right),
\end{split}
 \label{eq: interaction Hamiltonians}
\end{equation}
where $S_-^R=\dyad{\uparrow}{\Uparrow}$ and $S_-^L=\dyad{\downarrow}{\Downarrow}$, while  
$\kappa_k$ and $g$ denote coupling constants.

\subsection{Output field}

The derivation of the output field in terms of the spin system proceeds analogously for both polarizations, and as such 
we only present the essential steps for right-handed ($R$) polarization. 
In a standing-wave basis the electric field can be written as 
$\bm{E}_{R} = \bm{\hat{e}}_{R} E_R^{(+)}(z,t) + \bm{\hat{e}}^*_{R} E_R^{(-)}(z,t)$ with positive frequency component 

\begin{equation}
E_R^{(+)}(z,t) = \sum_k \sqrt{\frac{\omega_k}{4 \epsilon_0 A L}} \Big( \mathrm{e}^{i \phi_R} \mathrm{e}^{i \omega_k z/c} 
+ \mathrm{e}^{-i \omega_k z / c} \Big) r_{k,R}(t),
 \label{eq: field general}
\end{equation}
for $z > 0$. We write the Jones vectors for right/left handed circularly polarized light as $\bm{\hat{e}}_{R/L} = 1/\sqrt{2} (1,\pm i)$, respectively, and will also use $\bm{\hat{e}}_{H} = 1/\sqrt{2}(\bm{\hat{e}}_{R} + \bm{\hat{e}}_{L})$ and $\bm{\hat{e}}_{V} = -i/\sqrt{2}(\bm{\hat{e}}_{R} - \bm{\hat{e}}_{L})$ for horizontally and vertically polarized light in the following. $AL$ is the mode volume of the imagined port mode cavity, and $\phi_R$ denotes the phase incurred upon reflection from the top mirror.

Following a procedure detailed in Appendix \ref{app: output field} we can decompose the field operator $E_R^{(+)}(z,t)$ into the free field $E_{R,f}^{(+)}(z,t) \sim r_{k, R}(0) \exp(-i \omega_k t) \equiv r_{k, R}^f (t)$ reflected by the top mirror of the cavity, and the source field $E_{R,s}^{(+)}(z,t) \sim a_R(t)$, i.e. the field inside the cavity leaking through the top mirror into to port modes. This cavity field depends on the field of the dipole $\propto S_-^R$ due to the coupling Hamiltonian $H_{sc}^I$. We then assume that the dipole $S_-^R(t)$ undergoes dynamics with a Fourier transform centered at some frequency $\omega_d$ and a bandwidth much smaller than the cavity bandwidth $\kappa = 2 \pi g(\omega_c)|\kappa(\omega_c)|^2$, where $g(\omega_c)$ is the density of port modes at the cavity resonance frequency $\omega_c$. This assumption will be confirmed self-consistently in the Master equation treatment presented in Subsec. \ref{subsec: master equation}, where we find $\omega_d$ to correspond to the input laser frequency, while the bandwidth of dipole dynamics is limited by the decay rate $\Gamma \sim g^2/\kappa$. Working in the Purcell regime given by $\Gamma \ll \kappa$ we can then decompose the outgoing field evolving with retarded time $t_o = t-z/c$ as
\begin{equation}
E^{(+)}_{R,\mathrm{out}}(z,t) = \sum_k \sqrt{\frac{\omega_k}{4 \epsilon_0 A L}} R_k r_{k,R}^f(t_o) + \xi S_-^R(t_o),
 \label{eq: signal field}
\end{equation}
with $\xi = -2 g \frac{\kappa}{\kappa_c^*} \sqrt{\omega_c / (4 \epsilon_0 A L)} (i(\omega_c - \omega_d) + \kappa/2)^{-1}$ and where the reflection coefficient of the empty cavity is given by
\begin{equation}
R_k = \frac{i(\omega_c - \omega_k) - \kappa/2}{i(\omega_c - \omega_k) + \kappa/2}.
\end{equation}
The first term in Eq.~(\ref{eq: signal field}) can be interpreted as the light 
reflected by the cavity without having interacted with the electronic system, while the second term $\propto S_-^R (t_o)$ describes the light that is radiated by the dipole into the cavity mode and then transmitted through the top mirror into the output field. Note that solving the Heisenberg equation $\dot{S}_- (t)$ by Fourier transform in the weak-driving limit where $P_e \approx 0$ yields the phase shift obtained using input-output theory for a two-level system coupling to a cavity \cite{auffeves-garnier_giant_2007}, as shown in Appendix \ref{app: output field}.

While the selection rules single out the circular polarizations $R$ and $L$ as good bases for calculations involving the coupling Hamiltonian $H_{sc}^I$, an interesting experimental configuration is given by a horizontally polarized excitation laser and measurement of the scattered light in the orthogonal linear polarization. The corresponing fields are readily 
calculated from Eq.~(\ref{eq: signal field}), for instance the 
positive frequency component of the vertically polarized ($V$) field is given by
\begin{equation}\begin{split}
E_V^{(+)}(t) &= \frac{i}{\sqrt{2}}\left( E^{(+)}_{R,\mathrm{out}}(t) - E^{(+)}_{L,\mathrm{out}}(t) \right) \\
&= \sum_k \sqrt{\frac{\omega_k}{4 \epsilon_0 A L}} R_k r_{k,V}^f(t_o) + \xi S_-^V(t_o)
\end{split}
 \label{eq: vertically polarized output}
\end{equation}
The operators corresponding to vertical polarization are $r_{k,V} = i/\sqrt{2}\left( r_{k,R} - r_{k,L} \right)$ and $S_-^{V}(t) = i/\sqrt{2}\left( S_-^{R}(t) - S_-^{L}(t) \right)$. We can already see from this expression that if the input field is horizontally polarized, i.e. $\expect{r_{k,R}(0)} = \expect{r_{k,L}(0)}$, then $\expect{r_{k,V}^f(t_o)} = 0$ and therefore the vertically polarized output field (referred to as cross-polarized field) is solely dependent on the dipole field $\propto S_-^V$. In this way the excitation laser can be filtered out and we only need to consider the dipole field in the description of the cross-polarized field. We furthermore find that for a resonant and weak horizontally polarized input field all the laser power is scattered into the vertical polarization, in accordance with the input-output calculation used to explain the giant phase shift measured in \cite{androvitsaneas_efficient_2019}.

\subsection{Master equation for the spin system}
\label{subsec: master equation}
To calculate field expectation values using Eq.~(\ref{eq: signal field}) we need to find the time evolution of the 
dipole operator $S_-^R(t)$, and in order to do so we derive a master equation for the spin state 
$\rho$ in the Hilbert space spanned by $\{ \ket{\uparrow}, \ket{\Uparrow}, \ket{\downarrow}, \ket{\Downarrow} \}$. 
The derivation proceeds in two steps, the first tracing over the port modes to yield a 
master equation for the joint cavity mode--spin system, and the second step eliminates the cavity mode in the weak coupling Purcell regime. We assume the initial state $\ket{\psi_0}$ of the port mode system to be the vacuum except for two modes $r_{d,R}$ and $r_{d,L}$ satisfying $r_{d,R} \ket{\psi_0} = \alpha_{R/L} \ket{\psi_0}$, i.e. a coherent state describing a laser field oscillating at frequency $\omega_k = \omega_d$. Once again, the derivation for the two polarisation proceeds analogously. We refer to Appendix~\ref{app: Adiabatic elimination of the cavity mode} for the detailed derivation and state the resulting 
master equation for the spin system with adiabatically eliminated cavity mode in a frame rotating at $\omega_d$ as
\begin{equation}
\dot{\rho}(t) = -i [H_{s}, \rho(t)] + \sum_{k = R, L} L_k \rho (t) L_k^{\dagger} - \frac{1}{2} \left\{ L_k^{\dagger}L_k, \rho(t) \right\},
\label{eq: final Master equation}
\end{equation}
with
\begin{equation}
H_{s}= \Delta P_e + \omega_B \sigma_X + \left(\Omega_R S_-^R + \Omega_L S_-^L + \mathrm{H.c.} \right)
\label{SystemHamiltonian}
\end{equation}
and Lindblad operator $L_{R/L} = \sqrt{\Gamma} S_-^{R/L}$, where the Purcell enhanced decay rate and 
renormalised detuning are
\begin{equation}
\Gamma= \frac{g^2 \kappa}{\Delta_c^2  + \kappa^2/4} \ \mathrm{and} \ \Delta = \Delta_0 + \frac{\Delta_c g^2}{\Delta_c^2 + \kappa^2 / 4},
 \label{eq: decay rate and Lamb shift}
\end{equation}
where $\Delta_{0/c} = \omega_{0/c} - \omega_d$. The driving parameters are given by
\begin{equation}
\Omega_{R/L} = i \frac{g \alpha_{R/L}}{\kappa/2 - i(\omega_c - \omega_d)}.
\end{equation}
The spin system therefore obeys a master equation describing classical driving with Rabi frequencies $\Omega_{R/L}$ and a cavity-mediated decay rate $\Gamma$. Note that the polarization of the laser is encoded in the relative magnitude and phase of $\Omega_R$ and $\Omega_L$.

While the spin system basis spanned by $\left\{ \ket{\uparrow}, \ket{\Uparrow}, \ket{\downarrow}, \ket{\Downarrow} \right\}$ allowed us 
to treat polarizations $R$ and $L$ separately in the previous derivations, 
the so-called Voigt basis can be advantageous in the analysis of the dynamics of this spin system. 
The Voigt basis is $\{ \ket{+}, \ket{P}, \ket{-}, \ket{M} \}$ with basis states 

\begin{equation}
\begin{split}
\ket{\pm} &= \frac{1}{\sqrt{2}}\left( \ket{\uparrow} \pm \ket{\downarrow} \right), \\
\ket{P / M} &= \frac{1}{\sqrt{2}}\left( \ket{\Uparrow} \pm \ket{\Downarrow} \right).
\end{split}
\label{eq: Voigt states definition}
\end{equation}

In this basis the system Hamiltonian for a horizontally polarized laser ($\Omega_R = \Omega_L \equiv \Omega$) is given by (c.f. Eq.~({\ref{SystemHamiltonian}}))

\begin{equation}
H_{s} =
\begin{pmatrix}
\Delta & \Omega & 0 & 0 \\
\Omega & \omega_B & 0 & 0  \\
0 & 0 & - \omega_B & \Omega \\
0 & 0 & \Omega & \Delta
\end{pmatrix},
 \label{eq: Voigt basis Hamiltonian}
\end{equation}

while the spontaneous decay of the excited states is described by two Lindblad operators

\begin{equation}
\begin{split}
L_H &= \sqrt{\frac{\Gamma}{2}} \left( \dyad{+}{P} + \dyad{-}{M} \right) \equiv \sqrt{\Gamma} S^{H}_{-} \\
L_V &= \sqrt{\frac{\Gamma}{2}} \left( \dyad{+}{M} + \dyad{-}{P} \right) \equiv \sqrt{\Gamma} S^{V}_{-}.
\end{split}
 \label{eq: Voigt Lindblad operators}
\end{equation}

\begin{figure}
 \includegraphics[width = 0.6 \linewidth]{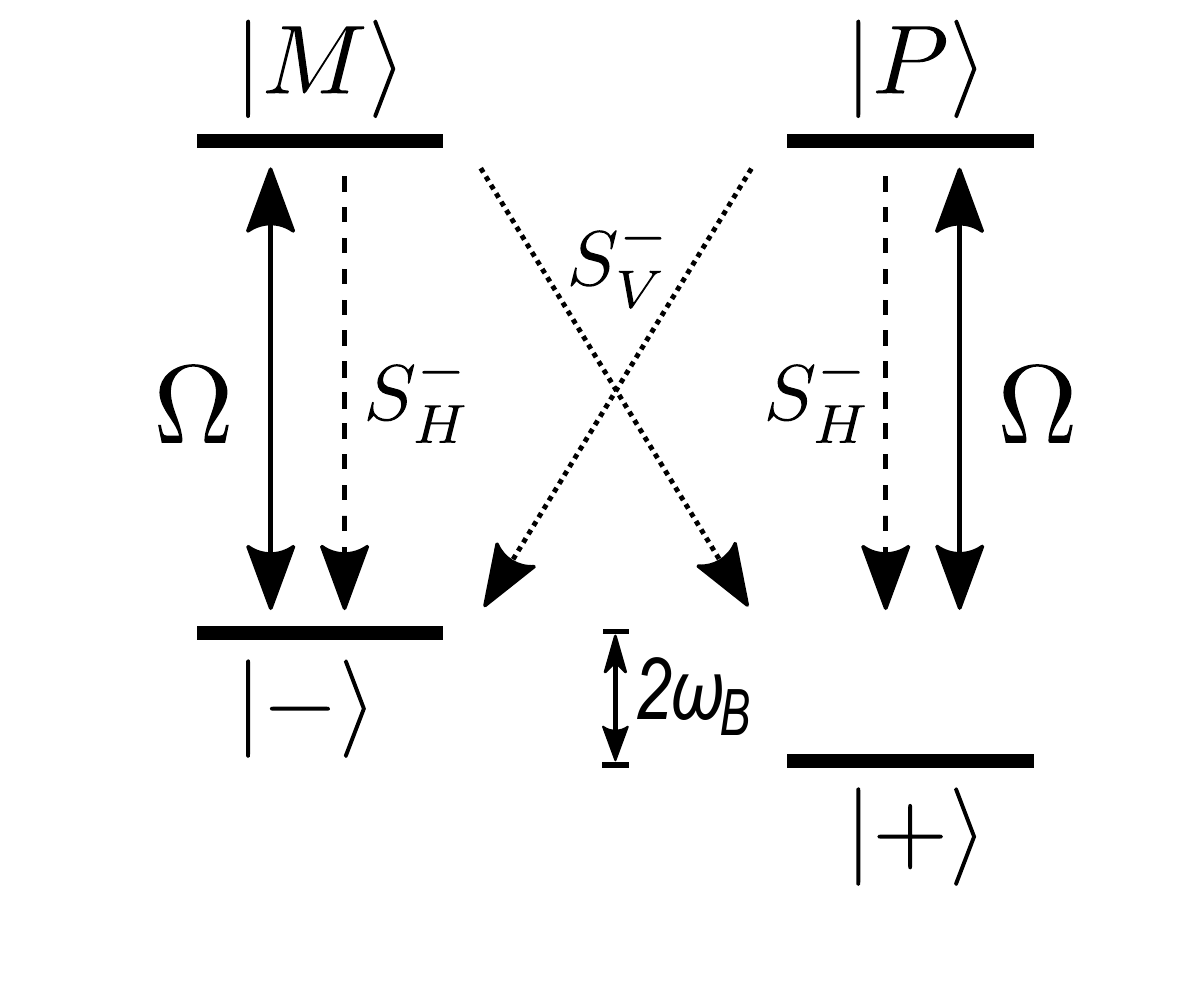}
 \caption{Level scheme in the Voigt basis for horizontally polarized driving. While the Hamiltonian does not couple the subsystems corresponding to right/left handed polarization any more, there is a decay term $\propto S^{V}_{-}$ that does introduce such transitions.}
 \label{fig: Voigt level scheme}
\end{figure}

Note that unlike in the Faraday basis, the Hamiltonian is block diagonal in the Voigt basis, while the decay operators are not. Which means that in this basis the the ground states are static within the magnetic field and population exchange is mediated through the optical transition which are excited by the drive laser. The energy level scheme and its couplings in this basis are shown in Fig. \ref{fig: Voigt level scheme}.

In this basis it is straight forward to apply the methods described in Ref.~\cite{reiter_effective_2012} to adiabatically eliminate the excited states (weak excitation approximation) which yields an effective master equation for the approximate state $\rho_{\mathrm{eff}}$ which exists in the ground state subspace spanned by $\{ \ket{+} ,\ket{-} \}$ alone. 

\begin{equation}
\dot{\rho}_{\mathrm{eff}}=-i[H_{\mathrm{eff}},\rho_{\mathrm{eff}}]\!+\!
\sum L_{\mathrm{eff}}^k\rho_{\mathrm{eff}}L_{\mathrm{eff}}^{k\dagger}-\frac{1}{2}\{L_{\mathrm{eff}}^{k\dagger}L_{\mathrm{eff}}^{k},\rho_{\mathrm{eff}}\}
\label{EffectiveME}
\end{equation}
where we have defined
\begin{equation}
\begin{split}
H_{\mathrm{eff}} &= \tilde{\omega}_B \big( \dyad{+}{+} - \dyad{-}{-} \big), \\
L_{\mathrm{eff}}^H &= \gamma_- \dyad{+}{+} + \gamma_+ \dyad{-}{-}, \\
L_{\mathrm{eff}}^V &= \gamma_- \dyad{-}{+} + \gamma_+ \dyad{+}{-},
\end{split}
 \label{eq: coherence diff eq}
\end{equation}
with renormalised Zeeman energy

\begin{equation}
\tilde{\omega}_B = \omega_B + \frac{1}{2}\left( \frac{\Omega^2 (\omega_B - \Delta)}{(\omega_B - \Delta)^2 + \left(\frac{\Gamma}{2}\right)^2} + \frac{\Omega^2 (\omega_B + \Delta)}{(\omega_B + \Delta)^2 + \left(\frac{\Gamma}{2}\right)^2} \right)
\end{equation}
and complex rates 

\begin{equation}
\gamma_- = \frac{\Omega \sqrt{\frac{\Gamma}{2}}}{\Delta - \omega_B - i \frac{\Gamma}{2}},
\qquad
\gamma_+ = \frac{\Omega \sqrt{\frac{\Gamma}{2}}}{\Delta + \omega_B - i \frac{\Gamma}{2}}.
\end{equation}
Inspection of this effective master equation reveals that optical driving directly affects the ground state spin dynamics.  There is a threefold effect of the optical driving, namely an energy shift described by $H_{\mathrm{eff}}$, a dephasing process described by $L_{\mathrm{eff}}^H$, 
and a spin flip process via virtual optical excitation corresponding to vertical decay processes described by $L_{\mathrm{eff}}^V$.

To understand the dynamics described by this effective master equation 
is it convenient to recast it in terms of a Bloch vector $\bm{r}$ which describes the quantum state of the spin $1/2$ particle 
in the ground state manifold spanned by $\{\ket{\pm}\}$. That is, we write 
$\rho_{\mathrm{eff}} = \frac{\mathbbm{1}}{2} + r_x \sigma_x + r_y \sigma_y + r_z \sigma_z$ where we define the Pauli matrices in the Voigt basis as $\sigma_z = \frac{1}{2} \left( \dyad{+}{+} - \dyad{-}{-} \right)$, $\sigma_x = \frac{1}{2} \left( \dyad{+}{-} + \dyad{-}{+}\right)$, $\sigma_y = i[\sigma_x, \sigma_z]$ and Bloch vector components $r_i =  \mathrm{Tr}\{\sigma_i \rho_{\mathrm{eff}}\}$.
We then find the components $r_x$, $r_y$ to obey 

\begin{equation}
\frac{d}{dt} \begin{pmatrix} r_x \\ r_y \end{pmatrix} = 
\begin{pmatrix} R -\gamma_{\Sigma} & -(2 \tilde{\omega}_B - I) \\ 2 \tilde{\omega}_B & -\gamma_{\Sigma} \end{pmatrix} \begin{pmatrix} r_x \\ r_y \end{pmatrix},
 \label{eq: equatorial Lindblad matrified}
\end{equation}

Where $\gamma_{\Sigma} = \abs{\gamma_+}^{2} + \abs{\gamma_-}^{2}$ and further $R$ and $I$ are defined through the following decomposition of $\gamma_+\gamma_-^{*} = \frac{R}{2} + i\frac{I}{2}$. This is a system of linear differential equations and can be solved by standard methods. We find the following solution. 
\begin{equation}
    \begin{pmatrix} r_x \\ r_y \end{pmatrix} = e^{-\gamma t} (c_{+} \textbf{v}_{+} e^{2i\omega_{e} t} + c_{-} \textbf{v}_{-} e^{-2i\omega_{e} t})
    \label{eq: coherence solutions}
\end{equation}
with total dephasing rate $\gamma = \gamma_{\Sigma} - R/2$. $\omega_{e}$ is given by 
\begin{equation}
    \omega_{e} = \sqrt{\tilde{\omega}_B(\tilde{\omega}_B - I/2) - (R/4)^{2}}.
    \label{eq: omegaE}
\end{equation}
This effective precession frequency depends on the renormalised Zeeman energy and the parameters R and I, this term characterises the competition between Larmor precession and dephasing processes within the ground state manifold. Finally, the vectors $\textbf{v}_{\pm}$ are given by 
\begin{equation}
    \textbf{v}_{\pm} = \begin{pmatrix} \frac{2(\tilde{\omega}_B - I/2)}{R/2 \mp 2i\omega_{e}} \\ 1 \end{pmatrix} = \begin{pmatrix} \mathrm{v}_{\pm} \\ 1 \end{pmatrix},
\end{equation}
and the coefficients $c_{\pm}$ are determined through initial conditions. Eq. \ref{eq: coherence solutions} shows the spin dynamics are described by damped harmonic motion, with frequency $2\omega_{e}$. 

\subsection{Derivation of optical coherence properties}
With our master equations in hand we can calculate the coherence properties of the scattered field. The properties of interest are captured by the steady state first order correlation function $g^{(1)}(\tau) = \langle E^{(-)}(\tau) E^{(+)}(0)\rangle$ which, using Eq.\ref{eq: vertically polarized output}, can be show to be proportional to $ \expect{S^{V}_{+}(\tau)S^{V}_{-}(0)}$. Making use of the quantum regression formula~\cite{carmichael1999statistical,mccutcheon2015optical} this is given by
\begin{equation}
\expect{S^{V}_{+}(\tau)S^{V}_{-}(0)}  = \mathrm{e}^{-i\omega_d \tau} \mathrm{Tr} \left\{ \bar{S}_{+}^{V} \exp (\mathcal{L} \tau) [\bar{S}_{-}^{V} \rho_{\mathrm{std}}] \right\},
 \label{eq: regression spectrum kernel}
\end{equation} 
where $\bar{S}^{V}_{-} = \exp(i\omega_d \tau) S^{V}_{-}$ is the dipole operator in the rotating frame, $\mathcal{L}$ denotes the Liovillian superoperator defined by the right hand side of Eq.~({\ref{eq: final Master equation}}), 
and $\rho_{\mathrm{std}}$ is the steady-state density operator of the same equation. 

We now calculate the steady state first order correlation function in the weak-excitation limit where $\Omega/\Gamma \ll 1$. Our procedure is to define an auxiliary density operator $\rho^{\prime}(\tau)$ which obeys $\dot{\rho}^{\prime}(\tau) = \mathcal{L} \rho^{\prime}(\tau)$ with $\rho^{\prime}(0) = \bar{S}_-^V \rho_{\mathrm{std}}$, from which the relevant correlation function can be found from $\mathrm{Tr} \left\{ \bar{S}_+^V \rho^{\prime}(\tau) \right\} = \bracket{-}{\rho^{\prime}(\tau)}{P} + \bracket{+}{\rho^{\prime}(\tau)}{M} \equiv \rho^{\prime}_{mP} + \rho^{\prime}_{pM}$. The relevant matrix elements of $\dot{\rho}^{\prime}(\tau)$ are

\begin{equation}\begin{split}
\frac{d}{d \tau} \rho^{\prime}_{mP} &= \left( i(\Delta + \omega_B) - \frac{\Gamma}{2} \right) \rho^{\prime}_{mP} - i \Omega \left(\rho^{\prime}_{MP} - \rho^{\prime}_{mp} \right) \\
\frac{d}{d \tau} \rho^{\prime}_{pM} &= \left( i(\Delta - \omega_B) - \frac{\Gamma}{2} \right) \rho^{\prime}_{pM} - i \Omega \left(\rho^{\prime}_{PM} - \rho^{\prime}_{pm} \right),
\end{split}
 \label{eq: off-diagonal Lindblad equation}
\end{equation}
where $\rho^{\prime}_{PM} \equiv \bracket{P}{\rho^{\prime}(\tau)}{M}$, $\rho^{\prime}_{MP} \equiv \bracket{M}{\rho^{\prime}(\tau)}{P}$. We can distinguish three types of matrix elements in these differential equations. First there are elements $\rho^{\prime}_{mp/pm}$ living in the ground state subspace. These matrix elements obey the effective master equation of Eq. \ref{EffectiveME} solved by Eq. \ref{eq: coherence solutions} and decay on a long timescale $1/\gamma \sim \Gamma/\Omega^2$. Secondly there are the matrix elements $\rho^{\prime}_{mP/pM}$ that obey Eq. \ref{eq: off-diagonal Lindblad equation} and behave like dipoles, i.e. they are suppressed by a decay term $\sim \Gamma$ but rise up due to a source term $\sim \Omega \rho^{\prime}_{mp/pm}$, which yields a term $\propto \Omega/\Gamma \rho^{\prime}_{mp/pm}$ upon integration. The third type of matrix elements is given by terms $\rho_{PM/MP}$ living entirely in the excited state subspace. These terms behave like excitation probabilities upon integration, i.e. they are suppressed by another factor $\Omega/\Gamma$ relative to dipole-like terms. In the weak-excitation approximation we neglect the terms $\rho_{PM/MP}$ in the excited state subspace, such that the differential equations of Eq. \ref{eq: off-diagonal Lindblad equation} simplify and can be integrated using the effective master equation solutions of Eq. \ref{eq: coherence solutions} to yield

\begin{equation}\begin{split}
\expect{\bar{S}^{V}_{+}(\tau) \bar{S}^{V}_{-}(0)} =\ & s_+ \mathrm{e}^{(2i \omega_e - \gamma) \tau}
+s_- \mathrm{e}^{(-2i \omega_e - \gamma) \tau},
\end{split}
\label{eq: Correlation function solution}
\end{equation}

where we have furthermore neglected fast-decaying terms $\propto \exp (-\Gamma \tau)$. The coefficients $s_{\pm}$, can be determined using the initial condition $\rho^{\prime}(0) = \bar{S}_V^- \rho_{\mathrm{std}}$ 
and are found to be 

\begin{equation}
    s_{\pm} = \frac{2\Omega^{2}}{N} \frac{(\mathrm{v}_{\mp}\omega_{B}+i\Delta - \frac{\Gamma}{2})(\mathrm{v}_{\pm}\alpha_{\pm} -\omega_{B})}{(\mathrm{v}_{\mp} - \mathrm{v}_{\pm})(\alpha_{\pm}^{2} + \omega_{B}^{2})},
    \label{eq: dipole matrix elements solutions}
\end{equation}
In this expression we have defined $\alpha_{\pm} = \left(\frac{\Gamma}{2} - i\Delta -\gamma\right) \pm 2i\omega_{e} 
$ and $N = \Gamma^{2} + 4\Delta^{2} + 4\omega_{B}^{2} + 8\Omega^{2}$.

This correlation function captures the coherence of cross-polarized light scattered by two coupled two-level systems in the weak-driving regime and is a generalisation of the seminal analysis of resonance fluorescence of a two-level system (TLS)~\cite{mollow_power_1969}.

It is easier to visualise the properties of this correlation function in the frequency domain hence we calculate the emission spectrum using the Wiener-Khinchin theorem. It states that in the steady-state the emission spectrum is proportional to the fourier transform of the two time electric field correlation function.
\begin{equation}
S(\omega) = 2 \mathrm{Re}\left[ \int_{0}^{\infty} d\tau \exp [ - i \omega \tau] \expect{S^{V}_{+}(\tau)S^{V}_{-}(0)} \right].
 \label{eq: spectrum positive time Fourier}
\end{equation}

Hence, for general $\omega_B$ but weak excitation we find from Eq.~(\ref{eq: Correlation function solution}) 

\begin{equation}
S(\omega - \omega_d) = 2 \mathrm{Re} \Bigg[ \frac{s_+}{ i \left(\omega - 2 \omega_e \right) + \gamma} + \frac{s_-}{ i \left(\omega + 2 \omega_e \right) + \gamma} \Bigg].
 \label{eq: weak-driving spectrum}
\end{equation}

When the coupling parameter $\omega_B = 0$ our theory reproduces the TLS analysis. This is the well-known coherent scattering (equivalently Heitler) limit in which the correlation function becomes $\expect{S^{V}_{+}(\tau)S^{V}_{-}(0)} = 2 \Omega^2/N$ and the corresponding emission spectrum takes the form $S(\omega) \propto \delta (\omega - \omega_d)$. It should be noted that in any real system environmental noise, such as that caused by nuclear spins, will induce additional spin dephasing effects that will not map onto the TLS analysis. The form of this spectrum constitutes one of our core findings, and is markedly different from the conventional resonance fluorescence spectrum in the coherent scattering regime. We will explore its implications in the following section. Furthermore, it should be noted that the resulting spectrum of the cross-polarized field is independent of the polarisation of the input field and it does not rely on the alignment along the field.

\section{Coherence and Spectral properties of the cross-polarized field }

\begin{figure}
    \centering
    \includegraphics[width=\linewidth]{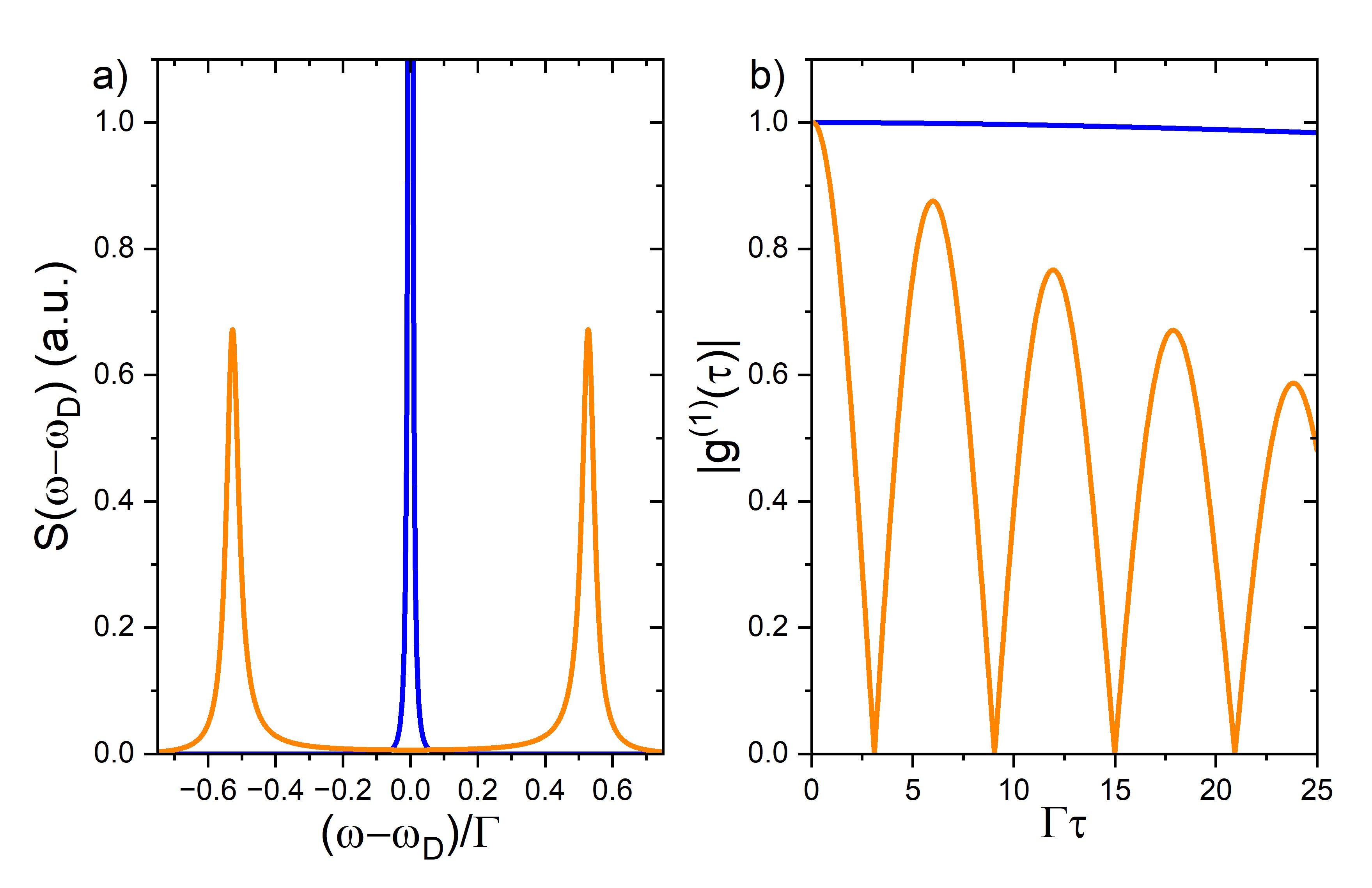}
    \caption{a) The blue curve shows the cross-polarised emission spectrum for the over damped regime, where depahsing induced by the drive field suppresses oscillations generated by the Larmor precession. The orange curve shows the cross-polarised emission spectrum for the under damped regime where the Larmor precession dominates. b) These plots show the absolute value of the first order correlation function for the same regimes (blue curve) over damped and (orange curve) under damped.}
    \label{fig:Dynamics}
\end{figure}
 
\begin{figure*}
    \includegraphics[width = \linewidth]{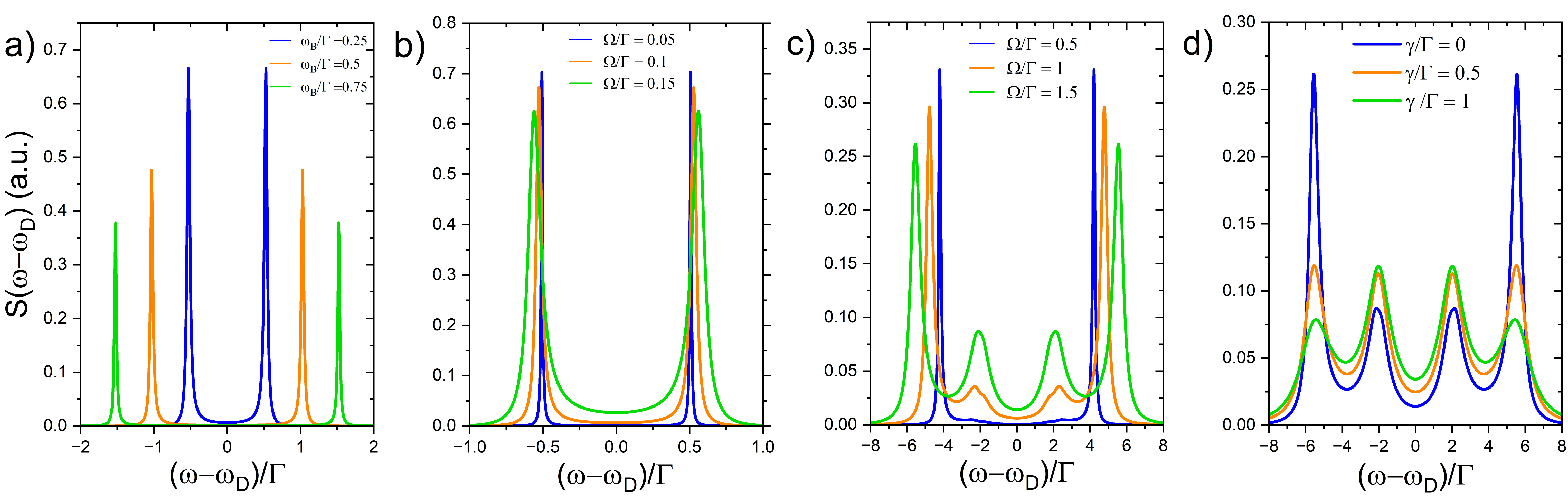}
    \caption{A selection of emission spectrum plots for different parameter regimes. a)The cross polarised emission spectrum for fixed driving power and varying magnetic field strength. This shows how the ground state coupling splits the spectrum by $4\omega_{B}$. b) The cross polarised emission spectrum for fixed magnetic field strength, $\omega_{B}/\Gamma = 0.25$, and varying Rabi frequency. These plots show that increased drive power leads to increase in the spectral broadening. This increase in damping also slightly modifies the natural precession frequency leading to a shift in the spectral positions. c) The cross polarised emission spectrum beyond the weak excitation approximation. Here we can see four peaks. The two outer peaks correspond to spin scattered light while the two inner peaks, centered at $\pm \omega_{B}$, are spontaneous emission from real excitation. d) The cross polarised emission spectrum with $\Omega/\Gamma = 1.5$ and varying spin dephasing rates. Showing that the outer spin scattered peaks are broadened by the ground state dynamics while the real excitation peaks are not.}
    \label{fig: Variety of Plots}
\end{figure*}
 In the previous section we showed that the spin dynamics are described by a damped harmonic system (Eq. \ref{eq: coherence solutions}), which will have regimes of qualitatively different behaviour, i.e., under/over damped. The novel observation here is that the source of damping derives from the optical driving of the system and scales with the strength of the drive field. The competition between this process and the Larmor precession induced by the applied magnetic field now define the regimes of harmonic motion for the spin, which is controlled by the parameter $\omega_{e}$. 
 
 When $\omega_{e}$ is a real number and $\gg 0$, i.e. large field strengths, the spin is under damped and  Eq.~(\ref{eq: coherence solutions}) 
describes Larmor precession with a frequency defined by the applied magnetic field ($2\omega_B$) and a damping rate $\gamma \sim {\Omega}^{2}/\Gamma$, related to the strength of the driving field. This behaviour is mapped to the output field and in the time domain gives rise to an oscillating first order correlation function (Eq. \ref{eq: Correlation function solution}) with an exponentially decaying envelope, this has been observed experimentally \cite{androvitsaneas2022quantum}. Transforming in to the frequency domain we find a split spectrum consisting of two lines at $\omega = \omega_d \pm 2 \omega_e$, 
both broadened by the optically induced spin dephasing rate $\gamma$, the orange curve in Fig. \ref{fig:Dynamics}(a). At the point where optical dephasing dominates the spin becomes over damped and $\omega_{e}$ become an imaginary number. The spin now dephases faster than the Larmor precession and Eq.~(\ref{eq: coherence solutions}) describes biexponential decay with rates $\gamma \pm 2|\omega_e|$. In the time domain of the output field the first order correlation function (Eq. \ref{eq: Correlation function solution}) no longer oscillates and simply decays, i.e., two overlapping spectral features both centred at $\omega = \omega_d$ of widths $\gamma \pm 2 |\omega_e|$, the blue curve in Fig. \ref{fig:Dynamics}(a). 

These spectral features constitute a novel and interesting regime of optical scattering. The position of the spectral lines is determined by the applied magnetic field and their width by the optical driving power. We see that the spectrum is split by approximately $4\omega_{B}$ (Fig. \ref{fig: Variety of Plots}(a)), i.e., twice the Zeeman energy. This is due to energy conservation, where production of vertically polarised light must coincide with a spin flip transition ($\ket{+} \Leftrightarrow \ket{-}$). This energy loss(gain) of $2\omega_{B}$ is compensated via a corresponding shift in the emitted photon to a higher(lower) energy. It is also clear from Fig. \ref{fig: Variety of Plots}(a) that increasing the magnetic field reduces the amount of light scattered into the vertically polarised output mode as the optical transitions detune from the input laser. Furthermore, in Fig. \ref{fig: Variety of Plots}(b) one observes that the spectral broadening of the individual peaks is controlled by the driving power. This gives rise to two sub natural linewidth peaks whose bandwidth is limited by the optically induced dephasing. This effect constitutes a new self-broadening mechanism of laser scattering that needs to be taken into account in optical studies of spin dephasing in the weak excitation regime. Taken together these unique features are not be captured via conventional treatment of resonance florescence from a TLS.

As the driving power is increased beyond the weak excitation limit our analytic solution for the spectrum begins to breakdown (see Appendix \ref{app: Breakdown}). To provide insight into the system's behavior beyond the weak excitation limit we can numerically solve Eq. \ref{eq: final Master equation}. This leads to Fig. \ref{fig: Variety of Plots}(c) showing that the outer spin scattered peaks are broadened but persist beyond the weak excitation regime. Additionally, the emergence of two other peaks centered at $\pm \omega_{B}$ is observed. These are due to real excitation giving rise to two spontaneous emission lines with bandwidth $\Gamma$. To verify that the outer peaks are indeed spin scattered light we add additional dynamics to the ground state subspace that model non optical environmental dephasing effects. To this end a term $L_{pd} = \sqrt{\frac{\gamma_{pd}}{2}} \sigma_{x}$ is added to the Master equation Eq. \ref{eq: final Master equation}. The emission spectra for different amounts of spin dephasing are shown in Fig. \ref{fig: Variety of Plots}(d). Clearly only the outer peaks are broadened by this dissipation term, further confirming the mapping of the spin dynamics to the output spectrum. 
\section{Conclusion}
We study an optically active spin $1/2$ system and find it to be an insightful, solvable model of coupled quantum emitters, thus generalizing the analysis of coherent scattering that has long served as a paradigm in quantum optics. Our analysis shows that in the weak driving limit the spectrum of spin scattered light is determined by the spin dynamics rather than the coherence of the excitation laser. This mapping could have useful implications for the measurement of quantum systems, particularly for the nuclear spin environments found in quantum dots, as numerous studies have shown that strong driving can lead to a complicated feedback mechanism between the electron and nuclear spin systems \cite{Economou-2014,Evers-2014}. Furthermore, the properties of the spin scattered field are important for understanding the application of these systems to quantum information tasks. For example, the efficiency/fidelity of protocols for generating entangled photons \cite{denning_protocol_2017-1,hu_giant_2008}, or implementing quantum memories or repeaters \cite{WaksDrop} can be sensitive to these spectral and coherence properties. The methods used in this analysis are also applicable to different configurations of coupled emitters, such as neutral or multiply charged quantum dots, defect centres or atoms. 
\bibliography{spinterferometer}

\begin{thebibliography}{10}
\providecommand{\url}[1]{#1}
\csname url@samestyle\endcsname
\providecommand{\newblock}{\relax}
\providecommand{\bibinfo}[2]{#2}
\providecommand{\BIBentrySTDinterwordspacing}{\spaceskip=0pt\relax}
\providecommand{\BIBentryALTinterwordstretchfactor}{4}
\providecommand{\BIBentryALTinterwordspacing}{\spaceskip=\fontdimen2\font plus
\BIBentryALTinterwordstretchfactor\fontdimen3\font minus
  \fontdimen4\font\relax}
\providecommand{\BIBforeignlanguage}[2]{{%
\expandafter\ifx\csname l@#1\endcsname\relax
\typeout{** WARNING: IEEEtran.bst: No hyphenation pattern has been}%
\typeout{** loaded for the language `#1'. Using the pattern for}%
\typeout{** the default language instead.}%
\else
\language=\csname l@#1\endcsname
\fi
#2}}
\providecommand{\BIBdecl}{\relax}
\BIBdecl

\bibitem{Humphreys-2018}
\BIBentryALTinterwordspacing
M.~J. e.~a. Humphreys~P.C., Kalb~N., ``Deterministic delivery of remote
  entanglement on a quantum network,'' \emph{Nature}, vol. 556, p. 268–273,
  June 2018. [Online]. Available:
  \url{https://www.nature.com/articles/s41586-018-0200-5}
\BIBentrySTDinterwordspacing

\bibitem{lindner_proposal_2009}
N.~H. Lindner and T.~Rudolph, ``\BIBforeignlanguage{en}{Proposal for {{Pulsed
  On}}-{{Demand Sources}} of {{Photonic Cluster State Strings}}},''
  \emph{\BIBforeignlanguage{en}{Physical Review Letters}}, vol. 103, no.~11, p.
  113602, Sep. 2009.

\bibitem{Gershoni-2023}
\BIBentryALTinterwordspacing
K.~O. e.~a. Cogan~D., Su~ZE., ``Deterministic generation of indistinguishable
  photons in a cluster state,'' \emph{Nat. Photon}, vol.~17, p. 324–329, Feb
  2023. [Online]. Available:
  \url{https://www.nature.com/articles/s41566-022-01152-2}
\BIBentrySTDinterwordspacing

\bibitem{economou_optically_2010}
S.~E. Economou, N.~Lindner, and T.~Rudolph, ``\BIBforeignlanguage{en}{Optically
  {{Generated}} 2-{{Dimensional Photonic Cluster State}} from {{Coupled Quantum
  Dots}}},'' \emph{\BIBforeignlanguage{en}{Physical Review Letters}}, vol. 105,
  no.~9, p. 093601, Aug. 2010.

\bibitem{lee_quantum_2019}
J.~P. Lee, B.~Villa, A.~J. Bennett, R.~M. Stevenson, D.~J.~P. Ellis, I.~Farrer,
  D.~A. Ritchie, and A.~J. Shields, ``\BIBforeignlanguage{en}{A quantum dot as
  a source of time-bin entangled multi-photon states},''
  \emph{\BIBforeignlanguage{en}{Quantum Sci. Technol.}}, vol.~4, no.~2, p.
  025011, Mar. 2019.

\bibitem{denning_protocol_2017-1}
E.~V. Denning, J.~{Iles-Smith}, D.~P.~S. McCutcheon, and J.~Mork, ``Protocol
  for generating multiphoton entangled states from quantum dots in the presence
  of nuclear spin fluctuations,'' \emph{Phys. Rev. A}, vol.~96, no.~6, p.
  062329, Dec. 2017.

\bibitem{schwartz_deterministic_2016}
I.~Schwartz, D.~Cogan, E.~R. Schmidgall, Y.~Don, L.~Gantz, O.~Kenneth, N.~H.
  Lindner, and D.~Gershoni, ``\BIBforeignlanguage{en}{Deterministic generation
  of a cluster state of entangled photons},''
  \emph{\BIBforeignlanguage{en}{Science}}, vol. 354, no. 6311, pp. 434--437,
  Oct. 2016.

\bibitem{vasconcelos_scalable_2019}
R.~Vasconcelos, S.~Reisenbauer, C.~Salter, G.~Wachter, D.~Wirtitsch,
  J.~Schmiedmayer, P.~Walther, and M.~Trupke,
  ``\BIBforeignlanguage{en}{Scalable spin-photon entanglement by
  time-to-polarization conversion},'' \emph{\BIBforeignlanguage{en}{arXiv
  preprint arXiv:1812.10338}}, Dec. 2019.

\bibitem{scerri_frequency-encoded_2018}
D.~Scerri, R.~N.~E. Malein, B.~D. Gerardot, and E.~M. Gauger,
  ``Frequency-encoded linear cluster states with coherent {{Raman}} photons,''
  \emph{Phys. Rev. A}, vol.~98, no.~2, p. 022318, Aug. 2018.

\bibitem{delteil_generation_2016}
A.~Delteil, Z.~Sun, W.-B. Gao, E.~Togan, S.~Faelt, and A.~Imamo{\u g}lu,
  ``Generation of heralded entanglement between distant hole spins,''
  \emph{Nature Physics}, vol.~12, pp. 218--223, Mar. 2016.

\bibitem{stockill_phase-tuned_2017}
R.~Stockill, M.~J. Stanley, L.~Huthmacher, E.~Clarke, M.~Hugues, A.~J. Miller,
  C.~Matthiesen, C.~Le~Gall, and M.~Atat{\"u}re, ``Phase-tuned entangled state
  generation between distant spin qubits,'' \emph{Phys. Rev. Lett.}, vol. 119,
  no.~1, p. 010503, Jul. 2017.

\bibitem{economou_unified_2005}
S.~E. Economou, R.-B. Liu, L.~J. Sham, and D.~G. Steel,
  ``\BIBforeignlanguage{en}{Unified theory of consequences of spontaneous
  emission in a {{$\Lambda$}} system},'' \emph{\BIBforeignlanguage{en}{Physical
  Review B}}, vol.~71, no.~19, p. 195327, May 2005.

\bibitem{nguyen_ultra-coherent_2011}
H.~S. Nguyen, G.~Sallen, C.~Voisin, P.~Roussignol, C.~Diederichs, and
  G.~Cassabois, ``Ultra-coherent single photon source,'' \emph{Applied Physics
  Letters}, vol.~99, no.~26, p. 261904, 2011.

\bibitem{hu_giant_2008}
C.~Y. Hu, A.~Young, J.~L. O'Brien, W.~J. Munro, and J.~G. Rarity,
  ``\BIBforeignlanguage{en}{Giant optical {{Faraday}} rotation induced by a
  single-electron spin in a quantum dot: {{Applications}} to entangling remote
  spins via a single photon},'' \emph{\BIBforeignlanguage{en}{Physical Review
  B}}, vol.~78, no.~8, p. 085307, Aug. 2008.

\bibitem{androvitsaneas_charged_2016}
P.~Androvitsaneas, A.~B. Young, C.~Schneider, S.~Maier, M.~Kamp, S.~Hoefling,
  S.~Knauer, E.~Harbord, C.~Y. Hu, J.~G. Rarity, and R.~Oulton,
  ``\BIBforeignlanguage{en}{Charged quantum dot micropillar system for
  deterministic light-matter interactions},''
  \emph{\BIBforeignlanguage{en}{Physical Review B}}, vol.~93, no.~24, p.
  241409, Jun. 2016.

\bibitem{hofmann_optimized_2003}
H.~F. Hofmann, K.~Kojima, S.~Takeuchi, and K.~Sasaki, ``Optimized phase
  switching using a single-atom nonlinearity,'' \emph{Journal of Optics B:
  Quantum and Semiclassical Optics}, vol.~5, no.~3, p. 218, 2003.

\bibitem{nutz_stabilization_2019}
T.~Nutz, P.~Androvitsaneas, A.~Young, R.~Oulton, and D.~P.~S. McCutcheon,
  ``\BIBforeignlanguage{en}{Stabilization of an optical transition energy via
  nuclear {{Zeno}} dynamics in quantum-dot\textendash{}cavity systems},''
  \emph{\BIBforeignlanguage{en}{Phys. Rev. A}}, vol.~99, no.~5, p. 053853, May
  2019.

\bibitem{mollow_power_1969}
B.~R. Mollow, ``Power spectrum of light scattered by two-level systems,''
  \emph{Phys. Rev.}, vol. 188, no.~5, p. 1969, Dec. 1969.

\bibitem{cresser1982resonance}
J.~Cresser, J.~H{\"a}ger, G.~Leuchs, M.~Rateike, and H.~Walther, ``Resonance
  fluorescence of atoms in strong monochromatic laser fields,'' in
  \emph{Dissipative Systems in Quantum Optics}.\hskip 1em plus 0.5em minus
  0.4em\relax {Springer}, 1982, pp. 21--59.

\bibitem{matthiesen_subnatural_2012}
C.~Matthiesen, A.~N. Vamivakas, and M.~Atat{\"u}re, ``Subnatural linewidth
  single photons from a quantum dot,'' \emph{Phys. Rev. Lett.}, vol. 108,
  no.~9, p. 093602, Feb. 2012.

\bibitem{proux_measuring_2015}
R.~Proux, M.~Maragkou, E.~Baudin, C.~Voisin, P.~Roussignol, and C.~Diederichs,
  ``Measuring the photon coalescence time window in the continuous-wave regime
  for resonantly driven semiconductor quantum dots,'' \emph{Phys. Rev. Lett.},
  vol. 114, no.~6, p. 067401, Feb. 2015.

\bibitem{bennett_cavity-enhanced_2016}
A.~J. Bennett, J.~P. Lee, D.~J.~P. Ellis, T.~Meany, E.~Murray, F.~F. Floether,
  J.~P. Griffths, I.~Farrer, D.~A. Ritchie, and A.~J. Shields,
  ``Cavity-enhanced coherent light scattering from a quantum dot,''
  \emph{Science Advances}, vol.~2, no.~4, p. 1501256, 2016.

\bibitem{warburton_single_2013}
R.~J. Warburton, ``Single spins in self-assembled quantum dots,'' \emph{Nature
  Materials}, vol.~12, no.~6, pp. 483--493, May 2013.

\bibitem{carmichael1999statistical}
H.~Carmichael, \emph{Statistical Methods in Quantum Optics 1: Master Equations
  and Fokker-Planck Equations}, ser. Theoretical and Mathematical
  Physics.\hskip 1em plus 0.5em minus 0.4em\relax {Springer}, 1999.

\bibitem{GIBBS197687}
H.~Gibbs and T.~Venkatesan, ``Direct observation of fluorescence narrower than
  the natural linewidth,'' \emph{Optics Communications}, vol.~17, no.~1, pp. 87
  -- 90, 1976.

\bibitem{volz_atom_2007}
J.~Volz, M.~Weber, D.~Schlenk, W.~Rosenfeld, C.~Kurtsiefer, and H.~Weinfurter,
  ``An atom and a photon,'' \emph{Laser Physics}, vol.~17, pp. 1007--1016, Jul.
  2007.

\bibitem{auffeves-garnier_giant_2007}
A.~{Auff{\`e}ves-Garnier}, C.~Simon, J.-M. G{\'e}rard, and J.-P. Poizat,
  ``\BIBforeignlanguage{en}{Giant optical nonlinearity induced by a single
  two-level system interacting with a cavity in the {{Purcell}} regime},''
  \emph{\BIBforeignlanguage{en}{Physical Review A}}, vol.~75, no.~5, p. 053823,
  May 2007.

\bibitem{androvitsaneas_efficient_2019}
P.~Androvitsaneas, A.~Young, J.~Lennon, C.~Schneider, S.~Maier, J.~Hinchliff,
  G.~Atkinson, E.~Harbord, M.~Kamp, S.~Hoefling, J.~G. Rarity, and R.~Oulton,
  ``\BIBforeignlanguage{en}{An efficient quantum photonic phase shift in a low
  {{Q}}-factor regime},'' \emph{\BIBforeignlanguage{en}{ACS Photonics}},
  vol.~6, no.~2, p. 8b01380, Jan. 2019.

\bibitem{reiter_effective_2012}
F.~Reiter and A.~S. S{\o}rensen, ``\BIBforeignlanguage{en}{Effective operator
  formalism for open quantum systems},'' \emph{\BIBforeignlanguage{en}{Physical
  Review A}}, vol.~85, no.~3, p. 032111, Mar. 2012.

\bibitem{mccutcheon2015optical}
D.~P.~S. McCutcheon, ``{Optical signatures of non-Markovian behavior in open
  quantum systems},'' \emph{Phys. Rev. A}, vol.~93, pp. 022\,119--022\,125,
  2016.

\bibitem{androvitsaneas2022quantum}
P.~Androvitsaneas, A.~B. Young, T.~Nutz, J.~M. Lennon, S.~Mister, C.~Schneider,
  M.~Kamp, S.~Höfling, D.~P.~S. McCutcheon, E.~Harbord, J.~G. Rarity, and
  R.~Oulton, ``Quantum modulation of a coherent state wavepacket with a single
  electron spin,'' 2022.

\bibitem{Economou-2014}
\BIBentryALTinterwordspacing
S.~E. Economou and E.~Barnes, ``Theory of dynamic nuclear polarization and
  feedback in quantum dots,'' \emph{Phys. Rev. B}, vol.~89, p. 165301, Apr
  2014. [Online]. Available:
  \url{https://link.aps.org/doi/10.1103/PhysRevB.89.165301}
\BIBentrySTDinterwordspacing

\bibitem{Evers-2014}
\BIBentryALTinterwordspacing
Y.~I. e.~a. Evers~E., Kopteva~N.E., ``Suppression of nuclear spin fluctuations
  in an ingaas quantum dot ensemble by ghz-pulsed optical excitation,''
  \emph{npj Quantum Inf}, vol.~7, Apr 2021. [Online]. Available:
  \url{https://www.nature.com/articles/s41534-021-00395-1}
\BIBentrySTDinterwordspacing

\bibitem{WaksDrop}
\BIBentryALTinterwordspacing
E.~Waks and J.~Vuckovic, ``Dipole induced transparency in drop-filter
  cavity-waveguide systems,'' \emph{Phys. Rev. Lett.}, vol.~96, p. 153601, Apr
  2006. [Online]. Available:
  \url{https://link.aps.org/doi/10.1103/PhysRevLett.96.153601}
\BIBentrySTDinterwordspacing

\bibitem{carmichael2007statistical}
H.~Carmichael, \emph{Statistical Methods in Quantum Optics 2: Non-Classical
  Fields}, ser. Theoretical and Mathematical Physics.\hskip 1em plus 0.5em
  minus 0.4em\relax {Springer}, 2007.

\end{thebibliography}
\bibliographystyle{IEEEtran}
\onecolumngrid
\appendix

\section{Derivation of the output field}
\label{app: output field}

Measured quantities are calculated from the upwards-propagating electric field $\bm{E}(z,t) = \bm{E}_R(z,t) + \bm{E}_L(z,t)$, which we decompose into positive and negative frequency components as $\bm{E}_{R/L} = \bm{E}_{R/L}^{(+)}(z,t) + \bm{E}_{R/L}^{(-)}(z,t)$. In terms of mode operators we write
\begin{equation}
\bm{E}_R^{(+)}(z,t) = i \bm{\hat{e}}_R \sum_k \sqrt{\frac{\omega_k}{\epsilon_0 A L^{\prime}}} \frac{1}{2 i} \Big( \mathrm{e}^{i \phi_R} \mathrm{e}^{i \omega_k z/c} + \mathrm{e}^{-i \omega_k z / c} \Big) r_{k,R}(t)
 \label{eq: field general appen}
\end{equation}
for $z > 0$ (similarly for $\bm{E}^{(+)}_L(z,t)$). In this expression $\bm{\hat{e}}_R = \frac{1}{\sqrt{2}} \begin{pmatrix}
    1 \\
    i 
\end{pmatrix}$
is the Jones vector for right-handed circularly polarized light propagating along $z$, $AL^{\prime}$ is the mode volume of the imagined port mode cavity, and $\phi_R$ denotes the optical phase incurred upon reflection from the top mirror.

Solving the Heisenberg equation $\dot{r}_{k,R}(t) = -i \omega_k r_{k, R}(t) - i \kappa_k a_R (t)$ obtained from Eq. \ref{eq: full Hamiltonian} yields
\begin{equation}
r_{k,R}(t) = \mathrm{e}^{- i \omega_k t} r_{k,R}(0) - i \kappa_k \int_0^t dt^{\prime} \mathrm{e}^{- i \omega_k (t-t^{\prime})} a_R(t^{\prime}).
\label{eq: formal solution r_k}
\end{equation}
Substituting this formal solution into Eq. \ref{eq: field general appen}, converting $\sum_k \rightarrow \int_0^{\infty} d \omega g(\omega)$ with a mode density $g(\omega)$, and approximating $g(\omega) \sqrt{\frac{\omega}{\epsilon_0 A L^{\prime}}} \kappa(\omega)$ constant over the bandwidth of the slowly varying cavity mode $\tilde{a}_R(t) = \exp(i \omega_c t) a_R(t)$ (Markov approximation) yields $\bm{E}^{(+)}_R (z,t) = \bm{E}^{(+)}_{R,f} (z,t) + \bm{E}^{(+)}_{R,s} (z,t)$, where 
\begin{equation}
\bm{E}^{(+)}_{R,f} (z,t) = i \bm{\hat{e}}_R \sum_k \sqrt{\frac{\omega_k}{\epsilon_0 A L^{\prime}}} \frac{1}{2 i} \Big( \mathrm{e}^{i \phi_R} \mathrm{e}^{-i \omega_k(t - \frac{z}{c}) } + \mathrm{e}^{-i \omega_k(t + \frac{z}{c}) } \Big) r_{k,R}(0)
 \label{eq: free field}
\end{equation}
and
\begin{equation}
\bm{E}^{(+)}_{R,s} (z,t) = - i \bm{\hat{e}}_R \mathrm{e}^{i \phi_R} \frac{\kappa}{2 \kappa(\omega_c)^*} \sqrt{\frac{\omega_c}{\epsilon_0 A L^{\prime}}} a_R(t-z/c),
 \label{eq: source field}
\end{equation}
and where we have defined $\kappa \equiv 2 \pi g(\omega_c)|\kappa(\omega_c)|^2$.
In the next step we decompose the source field $\propto a_R(t-z/c)$ into the reflected free field and the field emitted by the dipole $S_-^R$ in the adiabatic approximation, which assumes that the cavity bandwidth is much greater than the bandwidth of the dipole dynamics. The validity of this approximation can be confirmed self-consistently later on when we derive a master equation for the electron dynamics. 

The Heisenberg equation for mode operator $a_R(t)$ reads
\begin{equation}
\dot{a}_R(t) = -i \omega_c a_R(t) - i \sum_k \kappa_k^* r_{k,R}(t) - i g S_-^{R}(t).
 \label{eq: Heisenberg mode operator}
\end{equation}
substituting the formal solution $r_{k, R}(t)$ from Eq. \ref{eq: formal solution r_k} yields
\begin{equation}
\dot{a}_R (t) = - \left( i \omega_c + \frac{\kappa}{2} \right)a_R(t) - i \sum_k \kappa_k^* r_{k,R}(0) \mathrm{e}^{-i \omega_k t} - i g S_-^R (t)
 \label{eq: cavity mode diff eq}
\end{equation}
where again the Markov approximation for the cavity mode has been made. Integrating $\dot{a}_R (t)$ yields
\begin{equation}
a_R(t) = -i \sum_k \kappa_k^* r_{k,R}(0) \frac{\mathrm{e}^{-i \omega_k t}}{i(\omega_c - \omega_k) + \kappa / 2} - i \frac{g}{i(\omega_c - \omega_d) + \kappa/2} S_-^R(t)
\label{eq: cavity mode solution}
\end{equation}
for $t \gg 1/\kappa$, where $\omega_d$ gives the centre frequency of the oscillating dipole $S_-^R(t)$ (shown to equal the frequency of a driving field later on). A measured signal is given by the upwards propagating component of $\bm{E}(z,t)$, which is given by the components of $\bm{E}_{R/L}(z,t)$ that evolve with the retarded time $t-z/c$. Collecting these terms from Eq. \ref{eq: free field}, \ref{eq: source field}, and substituting from Eq. \ref{eq: cavity mode solution}, yields a signal field
\begin{equation}
\bm{E}^{(+)}_{R, out}(z,t) = \frac{\bm{\hat{e}}_R \mathrm{e}^{i \phi_R}}{\sqrt{\epsilon_0 A L^{\prime}}} \left[ \frac{1}{2} \sum_k \mathrm{e}^{-i \omega_k (t-z/c)}r_{k,R}(0) \sqrt{\omega_k} \left[ 1 - \sqrt{\frac{\omega_c}{\omega_k}}\frac{\kappa_k^*}{\kappa_c^*} \frac{\kappa}{i(\omega_c - \omega_k) + \kappa / 2} \right] - \frac{\sqrt{\omega_c}}{2 \kappa_c^*}\frac{\kappa g S_-^R(t-z/c)}{i(\omega_c - \omega_d) + \kappa/2} \right].
 \label{eq: signal field almost}
\end{equation}
For a near-resonant input field, i.e. $\braket{r_{k,R}^{\dagger}(0)r_{k,R}(0)} \approx 0$ for $|\omega_c - \omega_k|\gg \kappa$, we can approximate $\sqrt{\frac{\omega_c}{\omega_k}} \approx \frac{\kappa_k^*}{\kappa_c^*} \approx 1$ and obtain
\begin{equation}
\bm{E}^{(+)}_{R, out}(z,t) = \frac{\bm{\hat{e}}_R \mathrm{e}^{i \phi_R}}{2} \left[ \sum_k \sqrt{\frac{\omega_k}{\epsilon_0 A L^{\prime}}} \mathrm{e}^{-i\omega_k (t - z/c)} r_{k,R}(0) \frac{i(\omega_c - \omega_k) - \kappa/2}{i(\omega_c - \omega_k) + \kappa/2} - \sqrt{\frac{\omega_c}{\epsilon_0 A L^{\prime}}}\frac{\kappa g}{\kappa_c^*} \frac{S_-^R(t-z/c)}{i(\omega_c - \omega_d) + \kappa/2} \right].
 \label{eq: signal field full}
\end{equation}
The first term in the brackets of Eq. \ref{eq: signal field} can be interpreted as the light reflected by the cavity without having interacted with the electronic system ($\frac{i(\omega_c - \omega_k) - \kappa/2}{i(\omega_c - \omega_k) + \kappa/2}$ is the corresponding phase shift). The second term $\propto g S_-^R (t-z/c)$ then gives the light that is radiated by the dipole into the cavity mode and then transmitted through the top mirror into the output field.

\begin{figure}
 \includegraphics[width=0.4\linewidth]{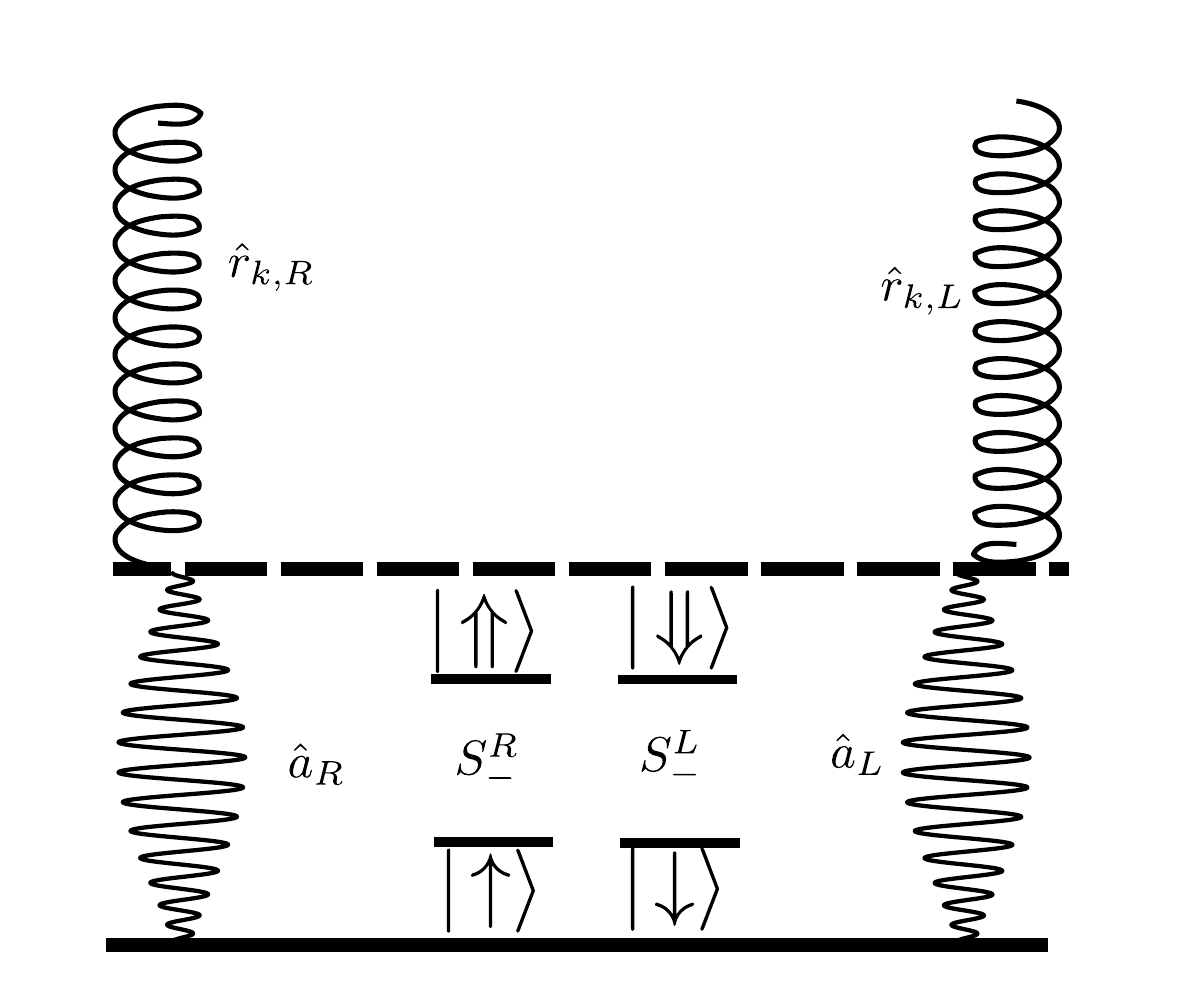}
 \caption{Components of the model. Showing the FLS coupled to the two cavity modes which in turn are coupled to the port modes.}
 \label{fig: diagram}
\end{figure}

\section{Master equation for the cavity-spin system}
\label{app: master equation for the cavity-spin system}
We eliminate the port modes $r_k$ from the full $\{ r_k, a, S_- \}$ system to obtain a Master equation for the driven, lossy cavity-spin subsystem described by $\{ a, S_- \}$ (we omit the labels $R$, $L$ in this section, as the derivation is the same for both polarizations). This derivation makes the assumption of a Markovian cavity, i.e. a cavity with a coupling parameter $\kappa_k$ that is essentially constant over its bandwidth.
We begin by splitting the Hamiltonian given in Eq. \ref{eq: free Hamiltonians} and Eq. \ref{eq: interaction Hamiltonians} into system, reservoir (i.e. port modes), and interaction components as $H = H_R^R + H_{SR}^R + H_R^L + H_{SR}^L + H_S$, where
\begin{equation}\begin{split}
H_R^{R/L} &=  \sum_k \omega_k r_{R/L}^{\dagger} r_{R/L} \\
H_{SR}^{R/L} &= \sum_k \left( \kappa_k^* r_{k, R/L} a^{\dagger} + \kappa_k r_{k, R/L}^{\dagger} a \right),
 \label{eq: system - reservoir port - cavity}
\end{split} \end{equation}
and where terms not containing port mode operators $r_k$ are contained in $H_S$. The following derivation applies to both circular polarizations $R$ and $L$ and we can omit these labels to yield a Hamiltonian $H = H_R + H_{SR} + H_S$, with
\begin{equation}
H_S = \omega_c a^{\dagger}a + \omega_0 P_e + g(S_+ a + S_- a^{\dagger}) + \omega_B \sigma_x,
 \label{eq: system Hamiltonian}
\end{equation}
where $P_e$ projects onto the excited state of a two-level system (TLS) (either $\{ \ket{\uparrow}, \ket{\Uparrow} \}$ or $\{ \ket{\downarrow}, \ket{\Downarrow} \}$) and $S_+$ and $S_-$ excite and de-excite this TLS, respectively. 

Starting with a state $\chi(t)$ of the combined reservoir-cavity-TLS system evolving under the above Hamiltonian, we define an interaction picture state $\tilde{\chi}(t) = \mathrm{e}^{i H_0 t} \chi(t) \mathrm{e}^{-i H_0 t}$ where $H_0 = H_R + \omega_d ( a^{\dagger}a + P_e)$. This state evolves as $\dot{\tilde{\chi}}(t) = - i [\tilde{H}(t), \tilde{\chi}(t)]$, with
\begin{equation}
\tilde{H}(t) = \sum_k  \left( \kappa_k^* r_k a^{\dagger} \mathrm{e}^{-i(\omega_k - \omega_d)t} + \kappa_k r_k^{\dagger} a \mathrm{e}^{i(\omega_k - \omega_d)t} \right) + H_S - \omega_d( a ^{\dagger}a + P_e).
 \label{eq: rotating frame Hamiltonian}
\end{equation}
We are interested in the reduced state of the cavity-TLS system $\tilde{\rho}(t) = \mathrm{Tr} \left\{ \tilde{\chi}(t) \right\}$, where the reservoir modes with operators $r_k$ are assumed to initially be in the vacuum state except for one mode $r_d$, which is in a coherent state $D_d \ket{0}$ due to the narrowband laser driving. To take this initial state into account we consider a state $\tilde{\chi}_D(t) \equiv D^{\dagger}_d \tilde{\chi}(t) D_d$, where $D_d$ is the displacement operator $D_d = \exp (\alpha r_d^{\dagger} - \alpha^* r_d)$ with dispacement amplitude $\alpha$. This displaced state $\tilde{\chi}_D(t)$ evolves as $\dot{\tilde{\chi}}_D(t) = -i [\tilde{H}_D(t) , \tilde{\chi}_D(t)]$, where $\tilde{H}_D(t) = D^{\dagger}_p \tilde{H}(t) D_d = \tilde{H}(t) + \kappa_d^* \alpha a^{\dagger} + \kappa_d \alpha^* a $, and importantly $\tilde{\rho}(t) = \mathrm{Tr} \left\{ \tilde{\chi}(t) \right\} = \mathrm{Tr} \left\{ \tilde{\chi}_D(t) \right\}$. We have therefore related a cavity interacting with a reservoir initially in a single-mode coherent state to a classically driven cavity interacting with a reservoir initially in the vacuum state. To remove the remaining system part of $\tilde{H}_D(t)$ we define another interaction picture by $\bar{\chi}_D(t) = \mathrm{e}^{i H_c t} \tilde{\chi}_D(t) \mathrm{e}^{-i H_c t}$ with $H_c = H_S - \omega_d( a ^{\dagger}a + P_e) + \kappa_d^* \alpha a^{\dagger} + \kappa_d \alpha^*a$ to find $\dot{\bar{\chi}}_D(t) = - i [\bar{H}_{SR}(t), \bar{\chi}_D(t)]$, with $\bar{H}_{SR}(t) = \bar{a}(t) \bar{\Gamma}^{\dagger}(t) + \bar{a}^{\dagger}(t) \bar{\Gamma}(t)$ and
\begin{equation}\begin{split}
\bar{a}(t) &= \mathrm{e}^{i H_c t} a \mathrm{e}^{-i H_c t} \mathrm{e}^{i \omega_d t} \\
\bar{\Gamma}(t) &= \sum_k \kappa_k^* r_k \mathrm{e}^{-i \omega_k t}.
 \label{eq: final interaction Hamiltonian}
\end{split} \end{equation}
Given this interaction Hamiltonian we can proceed with the derivation of a Master equation in the Born-Markov approximation following the steps given in detail in, for example \cite{carmichael1999statistical}.
Assuming an initially uncorrelated system-reservoir state $\bar{\chi}(0)$ we write
\begin{equation}
\dot{\bar{\rho}}(t) \approx - \int_0^t dt^{\prime} \mathrm{Tr} \left\{ \left[ \bar{H}_{SR}(t), \left[ \bar{H}_{SR}(t^{\prime}), \bar{\rho}(t^{\prime}) \otimes R_0 \right] \right] \right\},
 \label{eq: Born approximation}
\end{equation}
where the Born approximation consists of the replacement $\bar{\chi}_D(t^{\prime}) \rightarrow \bar{\rho}(t^{\prime})\otimes R_0$ where $R_0$ denotes the vacuum state of the reservoir. This approximation therefore assumes that the effect of changes to the reservoir on the system dynamcics due to its interaction with the system are negligible, which is justified in the case of the large reservoir considered here. More explicitly we write
\begin{equation} \begin{split}
\dot{\bar{\rho}}(t) = - \int_0^t dt^{\prime} \Big( & \bar{a}^{\dagger}(t)\bar{a}(t^{\prime}) \bar{\rho}(t^{\prime}) \mathrm{Tr} \left\{ \bar{\Gamma}(t) \bar{\Gamma}^{\dagger}(t^{\prime}) R_0 \right\} + \bar{a}(t) \bar{\rho}(t^{\prime})\bar{a}^{\dagger}(t^{\prime}) \mathrm{Tr} \left\{ \bar{\Gamma}^{\dagger}(t) R_0 \bar{\Gamma}(t^{\prime}) \right\} + \\
& \bar{a}(t^{\prime}) \bar{\rho}(t^{\prime})\bar{a}^{\dagger}(t) \mathrm{Tr} \left\{ \bar{\Gamma}^{\dagger}(t^{\prime}) R_0 \bar{\Gamma}(t) \right\} + \bar{\rho}(t^{\prime}) \bar{a}^{\dagger}(t^{\prime}) \bar{a}(t) \mathrm{Tr} \left\{ \bar{\Gamma}(t^{\prime}) \bar{\Gamma}^{\dagger}(t) R_0 \right\} \Big),
 \label{eq: Born explicit}
\end{split} \end{equation}
Evaluating
\begin{equation}
\mathrm{Tr} \left\{ \bar{\Gamma}(t) \bar{\Gamma}^{\dagger}(t^{\prime}) R_0 \right\} = \int_0^{\infty} d\omega g(\omega) |\kappa (\omega)|^2 \mathrm{e}^{i \omega (t^{\prime} - t)}
 \label{eq: port mode correlation function}
\end{equation}
with $g(\omega)$ denoting the density of states of the port modes, we arrive at four terms of the form
\begin{equation}
\int_0^t dt^{\prime} \bar{a}^{\dagger}(t)\bar{a}(t^{\prime}) \bar{\rho}(t^{\prime}) \mathrm{Tr} \left\{ \bar{\Gamma}(t) \bar{\Gamma}^{\dagger}(t^{\prime}) R_0 \right\} = \int_0^{\infty} d \omega g(\omega) |\kappa (\omega)|^2 \int_0^t dt^{\prime} \bar{a}^{\dagger}(t)\bar{a}(t^{\prime}) \bar{\rho}(t^{\prime})\mathrm{e}^{-i \omega (t - t^{\prime})}.
 \label{eq: just before Markov}
\end{equation}
The term $\bar{a}^{\dagger}(t)\bar{a}(t^{\prime}) \bar{\rho}(t^{\prime})$ varies slowly with $t^{\prime}$, as its evolution is generated by $\tilde{H}_D(t)$, which only features terms much smaller than the optical frequency $\omega_d$. The integral therefore averages to zero for frequencies $\omega$ that significantly differ from $\omega_c \approx \omega_0 \approx \omega_d$. Thus we can approximate $g(\omega) |\kappa (\omega)|^2 \approx g(\omega_c) |\kappa (\omega_c)|^2$ under the integral $\int_0^{\infty} d \omega$, extend the lower limit $0 \rightarrow - \infty$, evaluate $\int_{-\infty}^{\infty} d \omega \exp (-i \omega (t - t^{\prime})) = 2 \pi \delta (t - t^{\prime})$, and arrive at 
\begin{equation}
\dot{\bar{\rho}}(t) = \kappa \left[ \bar{a}(t) \bar{\rho}(t) \bar{a}^{\dagger}(t) - \frac{1}{2}\left( \bar{a}^{\dagger}(t)\bar{a}(t) \bar{\rho}(t) + \bar{\rho}(t) \bar{a}^{\dagger}(t)\bar{a}(t) \right) \right],
 \label{eq: cavity Master equation whacky frame}
\end{equation}
where $\kappa \equiv 2 \pi g(\omega_c) |\kappa (\omega_c)|^2$. Reversing the transformations $\bar{\rho}(t) \rightarrow \tilde{\rho}(t) \rightarrow \rho(t)$ and including the mode polarization labels yields the master equation for the cavity-TLS system in the Born-Markov approximation:
\begin{equation} \begin{split}
\dot{\rho}(t) &= -i [H, \rho(t)] + \sum_{k = R, L} \left( L_k \rho (t) L_k^{\dagger} - \frac{1}{2}\left( L_k^{\dagger}L_k \rho(t) + \rho(t) L_k^{\dagger}L_k \right) \right) \\
&\mathrm{where} \\
H &= \omega_c (a_R^{\dagger} a_R + a_L^{\dagger}a_L) + \epsilon_R \mathrm{e}^{i \omega_d t} a_R + \epsilon_R^* \mathrm{e}^{-i \omega_d t} a^{\dagger}_R + \epsilon_L \mathrm{e}^{i \omega_d t} a_L + \epsilon_L^* \mathrm{e}^{-i \omega_d t} a^{\dagger}_L \\
&\ \ \ + \omega_0 (P_{\Uparrow} + P_{\Downarrow}) + \omega_B \sigma_X + g \left(S_-^R a_R^{\dagger} + S_+^R a_R + S_-^L a_L^{\dagger} + S_+^L a_L \right), \\
L_R &= \sqrt{\kappa} a_R , \\
L_L &= \sqrt{\kappa} a_L.
 \label{eq: cavity Master equation finally}
\end{split} \end{equation}

\section{Adiabatic elimination of the cavity mode}
\label{app: Adiabatic elimination of the cavity mode}
The procedure presented in Appendix \ref{app: master equation for the cavity-spin system} has eliminated the cavity modes $r_{k, R/L}$ from the system description to yield a model in terms of two cavity modes $a_R$ and $a_L$ coupling to the electronic four-level system (FLS). In order to evaluate the output field given in Eq. \ref{eq: source field}, however, we only need the FLS dynamics, such that we can further simplify the analysis by eliminating the cavity modes from the description to arrive at a master equation of the FLS alone. This elimination is possible in the adiabatic limit, which assumes that the cavity linewidth $\sim \kappa$ is much greater than the bandwidth of the FLS dynamics. In physical terms this adiabatic approximation assumes that the cavity modes couple much stronger to the port modes than to the FLS dipoles. As far as the FLS dynamics are concerned these cavity modes can therefore be assumed in a state governed by the driving of the cavity rather than photons emitted by the FLS, which escape from the cavity so fast that they do not interact significantly with the FLS. Mathematically the derivation of the FLS Master equation is somewhat different from the previous calculation, as the starting point is the Master equation Eq. \ref{eq: cavity Master equation finally} rather than unitary evolution under the Hamiltonian defined in Eq. \ref{eq: free Hamiltonians} and \ref{eq: interaction Hamiltonians}. We make use of a superoperator formalism and closely follow \cite{carmichael2007statistical}.

Again the derivation proceeds independently and in the same way for the two polarizations $R$ and $L$, such that we can consider a simplified version of Eq. \ref{eq: cavity Master equation finally}, i.e. a cavity-TLS state $\rho(t)$ evolving as $\dot{\rho}(t) = -i [H, \rho(t)] + L \rho (t) L^{\dagger} - \frac{1}{2}\left( L^{\dagger}L \rho(t) + \rho(t) L^{\dagger}L \right)$ with
\begin{equation}\begin{split}
H &= \omega_c a^{\dagger} a + \omega_0 P_e + \epsilon \mathrm{e}^{i \omega_d t} a + \epsilon^* \mathrm{e}^{-i \omega_d t} a^{\dagger} + g (S_- a^{\dagger} + S_+ a ) \\
L &= \sqrt{\kappa} a .
\label{eq: simplified adiabatic starting point}
\end{split}\end{equation}
In a rotating frame $\tilde{\rho}(t) = \exp(i \omega_d (a^{\dagger} a + P_e)) \rho(t) \exp(-i \omega_d (a^{\dagger} a + P_e))$ the Hamiltonian becomes 
\begin{equation}
H = \Delta_c a^{\dagger} a + \Delta_0 P_e + \epsilon a + \epsilon^* a^{\dagger} + g (S_- a^{\dagger} + S_+ a )
\label{eq: rotating frame adiabatic}
\end{equation}
with $\Delta_{c/0} = \omega_{c/0} - \omega_d$.
We can again introduce a displacement operator in order to have a more convenient form of the Born approximation later on. Consider $\tilde{\rho}_D(t) = D(\beta)^{\dagger} \tilde{\rho}(t) D(\beta)$ with displacement operator $D(\beta) = \exp (\beta a^{\dagger} + \beta^* a)$. This displaced state evolves as
\begin{equation}
\dot{\tilde{\rho}}_D(t) = -i [H_D, \tilde{\rho}_D(t)] + \kappa \left( a_D \tilde{\rho}_D(t) a_D^{\dagger} - \frac{1}{2}\left( a_D^{\dagger} a_D \tilde{\rho}_D(t) + \tilde{\rho}_D(t) a_D^{\dagger} a_D \right) \right)
\label{eq: displaced master equation adiabatic before}
\end{equation}
where $a_D \equiv D(\beta)^{\dagger} a D(\beta) = a + \beta$. Choosing $\beta = -i \frac{\epsilon^*}{\kappa/2 + i \Delta_c}$ we find 
\begin{equation}
\dot{\tilde{\rho}}_D(t) = -i [H_D, \tilde{\rho}_D(t)] + \kappa \left( a \tilde{\rho}_D(t) a^{\dagger} - \frac{1}{2}\left( a^{\dagger} a \tilde{\rho}_D(t) + \tilde{\rho}_D(t) a^{\dagger} a \right) \right),
\label{eq: displaced master equation adiabatic}
\end{equation}
where $H_D = \Delta_c a^{\dagger}a + \Delta_0 P_e + g(a^{\dagger}S_- + a S_+) + \Omega S_- + \Omega^* S_+$ and where we define the Rabi frequency $\Omega = g \beta^*$. As in the derivation of the cavity master equation we have arrived at a classical driving Hamiltonian, where in this case it is the TLS rather than the cavity that is classically driven.

The following steps are conveniently expressed in a superoperator formalism following \cite{carmichael2007statistical}. We write $\mathcal{L} = (\ \bullet \ A)$ to denote an operation acting on an arbitrary operator $O$ as $\mathcal{L} A = O A$. The bullet $\bullet$ in the definition denotes whether the operator to be acted on is to be right-: $( \bullet \ A)$ or left-multiplied: $(A \ \bullet \ )$, or sandwiched in between: $(A \ \bullet \ B)$. We can then write Eq. \ref{eq: displaced master equation adiabatic} as $\dot{\tilde{\rho}}_D(t) = (\mathcal{L}_S + \mathcal{L}_R + \mathcal{L}_{SR})\tilde{\rho}_D(t)$, where
\begin{equation}\begin{split}
\mathcal{L}_S &= -i \Big( (H_S \ \bullet \ ) - (\ \bullet \ H_S) \Big) \\
\mathcal{L}_R &= -i \Delta_c \Big( (a^{\dagger} a \ \bullet \ ) - (\ \bullet \ a^{\dagger} a) \Big) + \kappa \Big( (a\ \bullet \ a^{\dagger}) - \frac{1}{2}(a^{\dagger}a \ \bullet \ ) - \frac{1}{2}(\ \bullet \ a^{\dagger}a) \Big) \\
\mathcal{L}_{SR} &= g \Big( (a^{\dagger} \ \bullet \ )(S_- \ \bullet \ ) - (a \ \bullet \ )(S_+ \ \bullet \ ) + (\ \bullet \ a)(\ \bullet \ S_+) - (\ \bullet \ a^{\dagger})(\ \bullet \ S_-) \Big).
 \label{eq: superoperator definitions}
\end{split}\end{equation}
All TLS terms have been subsumed in $H_S$, i.e. $H_D = \Delta_c a^{\dagger}a + H_S + g (a^{\dagger}S_- + a S_+)$.

Our goal is to derive a Master equation for the TLS state $\dot{\tilde{\sigma}}(t) = \mathrm{Tr}_R \left\{ \dot{\tilde{\rho}}(t) \right\} = \mathrm{Tr}_R \left\{ \dot{\tilde{\rho}}_D(t) \right\}$. We begin by defining an interaction picture as $\bar{\rho}(t) = \exp (-(\mathcal{L}_P + \mathcal{L}_S)t) \tilde{\rho}(t)$, where $\dot{\bar{\rho}}(t) = \bar{\mathcal{L}}_{SR} \bar{\rho}(t)$, $\bar{\mathcal{L}}_{SR} = \exp (-(\mathcal{L}_P + \mathcal{L}_S)t) \mathcal{L}_{SR}\exp ((\mathcal{L}_P + \mathcal{L}_S)t)$, and $\bar{\sigma}(t) = \exp (- \mathcal{L}_S) \tilde{\sigma}(t)$. Then we find
\begin{equation}
\dot{\bar{\sigma}}(t) = \int_0^t dt^{\prime} \mathrm{Tr}_R \left\{ \bar{\mathcal{L}}_{SR}(t) \bar{\mathcal{L}}_{SR}(t^{\prime}) \bar{\sigma}(t^{\prime})\otimes A_0 \right\}
 \label{eq: adiabatic elimination Born expression}
\end{equation}
given an uncorrelated initial state $\bar{\rho}(0)$ and making the Born approximation $\bar{\rho}(t^{\prime}) \rightarrow \bar{\sigma}(t^{\prime}) \otimes A_0$ under the integral $\int_0^t dt^{\prime}$, where $A_0$ denotes the vacuum state of the cavity mode obeying $a A_0 = A_0 a^{\dagger} = 0$. This Born approximation constitutes the first part of the adiabatic approximation, which assumes that the effect of the cavity mode on the TLS is well described by the interaction with a coherent state. We can then write explicitly
\begin{equation}
\mathcal{\bar{L}}_{SR} = - i g \Big( (\overline{a^{\dagger} \ \bullet \ })(\overline{S_- \ \bullet \ }) + (\overline{a \ \bullet \ \vphantom{+1}})(\overline{S_+ \ \bullet \ }) - (\overline{\ \bullet \ a \vphantom{+1}})(\overline{\ \bullet \ S_+}) - (\overline{\ \bullet \ a^{\dagger}})(\overline{\ \bullet \ S_-}) \Big)
 \label{eq: mod superoperator definitions}
\end{equation}
where $\{ (\overline{S_- \ \bullet \ }), (\overline{S_+ \ \bullet \ }), (\overline{\ \bullet \ S_+ }), (\overline{\ \bullet \ S_- }) \} = \exp (- \mathcal{L}_S t) \{ (S_- \ \bullet \ ), (S_+ \ \bullet \ ), (\ \bullet \ S_+ ), (\ \bullet \ S_- ) \} \exp (\mathcal{L}_S t)$ and similarly for reservoir superoperators such as $(\overline{\ \bullet \ a \vphantom{+1}}) = \exp (-\mathcal{L}_R t) (\ \bullet \ a) \exp (\mathcal{L}_R t)$. Following \cite{carmichael2007statistical} we evaluate
\begin{equation}
 \begin{split}
 (\overline{a\ \bullet \  \vphantom{+2}}) &= \exp \left(- \left(i \Delta_c + \frac{\kappa}{2}\right)t\right) (a \ \bullet \ ) \\
 (\overline{a^{\dagger} \ \bullet \  \vphantom{+2}}) &= \exp \left(\left(i \Delta_c + \frac{\kappa}{2}\right)t\right) (a^{\dagger} \ \bullet \ ) + \left[\exp \left(\left(i \Delta_c + \frac{\kappa}{2}\right)t\right) - \exp \left(-\left(i \Delta_c + \frac{\kappa}{2}\right)t\right) \right] ( \ \bullet \ a^{\dagger}).
 \end{split}
\label{eq: evaluated reservoir superoperators}
\end{equation}
Using these expressions as well as their conjugates, the integrand of Eq. \ref{eq: adiabatic elimination Born expression} becomes
\begin{equation}\begin{split}
\mathrm{Tr}_R & \left\{ \bar{\mathcal{L}}_{SP}(t) \bar{\mathcal{L}}_{SP}(t^{\prime}) \bar{\sigma}(t^{\prime})\otimes A_0 \right\} = \\
g^2 \Big[ & \exp \left(- \left(i \Delta_c + \frac{\kappa}{2}\right)(t - t^{\prime})\right) \left[ (\overline{\ \bullet \ S_+ })(t) (\overline{S_- \ \bullet \ })(t^{\prime})  - (\overline{S_+ \ \bullet \ })(t) (\overline{S_- \ \bullet \ })(t^{\prime}) \right] + \\
 & \exp \left(- \left(-i \Delta_c + \frac{\kappa}{2}\right)(t - t^{\prime})\right) \left[(\overline{S_- \ \bullet \ })(t) (\overline{\ \bullet \ S_+ })(t^{\prime}) - (\overline{\ \bullet \ S_- })(t) (\overline{\ \bullet \ S_+ })(t^{\prime}) \right] \Big] \bar{\sigma}(t^\prime).
 \label{eq: integrand adiabatic elimination}
\end{split} \end{equation}
The second part of the adiabatic approximation assumes that the cavity bandwidth $\kappa$ is much greater than the bandwidth of the system dynamics, which means for the above expression that $(\overline{\ \bullet \ S_+ })(t) (\overline{S_- \ \bullet \ })(t^{\prime}) \approx (\overline{\ \bullet \ S_+ })(t) (\overline{S_- \ \bullet \ })(t)$ for $t-t^{\prime} \in [0, 1/\kappa]$. We can therefore evaluate the integral $\int_0^t d t^{\prime}$ in Eq. \ref{eq: adiabatic elimination Born expression} to obtain
\begin{equation}\begin{split}
\dot{\bar{\sigma}}(t) = \frac{\kappa g^2}{2 \Delta_c^2 + \kappa^2/2} \left[ 2 (\overline{S_- \bullet S_+})(t) - (\overline{S_+ S_- \bullet })(t) - (\overline{\bullet S_+ S_- })(t) \right] \\
- \frac{i \Delta_c}{\Delta_c^2 + \kappa^2 / 4} \left[(\overline{S_+ S_- \bullet })(t) - (\overline{\bullet S_+ S_- })(t) \right].
 \label{eq: adiabatic elimination result interaction picture}
\end{split} \end{equation}
The first line of this expression describes the cavity-mediated decay rate of the TLS (Purcell effect), while the second line gives a Lamb shift of the TLS resonance energy. Undoing the interaction picture (but maintaining the frame rotating at $\omega_d$) and including both polarizations we therefore obtain the Master equation for the FLS with adiabatically eliminated cavity modes:
\begin{equation}\begin{split}
\dot{\tilde{\sigma}}(t) &= - i [H, \tilde{\sigma}(t)] + \sum_{k = R, L} \left( L_k \tilde{\sigma}(t) L_k^{\dagger} - \frac{1}{2}\left( \tilde{\sigma}(t)L_k^{\dagger} L_k +  L_k^{\dagger} L_k \tilde{\sigma}(t)\right) \right) \\
 \mathrm{where} \ 
H &= \Delta^{\prime}(P_{\Uparrow} + P_{\Downarrow}) +  \omega_B \sigma_X + \Omega_R S_-^R + \Omega_R^* S_+^R + \Omega_L S_-^L + \Omega_L^* S_+^L , \\
L_{R/L} &= \sqrt{\Gamma} S_-^{R/L}, \\ 
\mathrm{with}\  
\Gamma &= \frac{g^2 \kappa}{\Delta_c^2  + \kappa^2/4} \ \ \mathrm{and} \ \ \Delta = \left( \Delta_0 + \frac{\Delta_c g^2}{\Delta_c^2 + \kappa^2 / 4} \right).
 \label{eq: final Master equation appendix}
\end{split}\end{equation}

\section{Breakdown of Approximation}
\label{app: Breakdown}
\begin{figure*}[h]
\includegraphics[width = \linewidth]{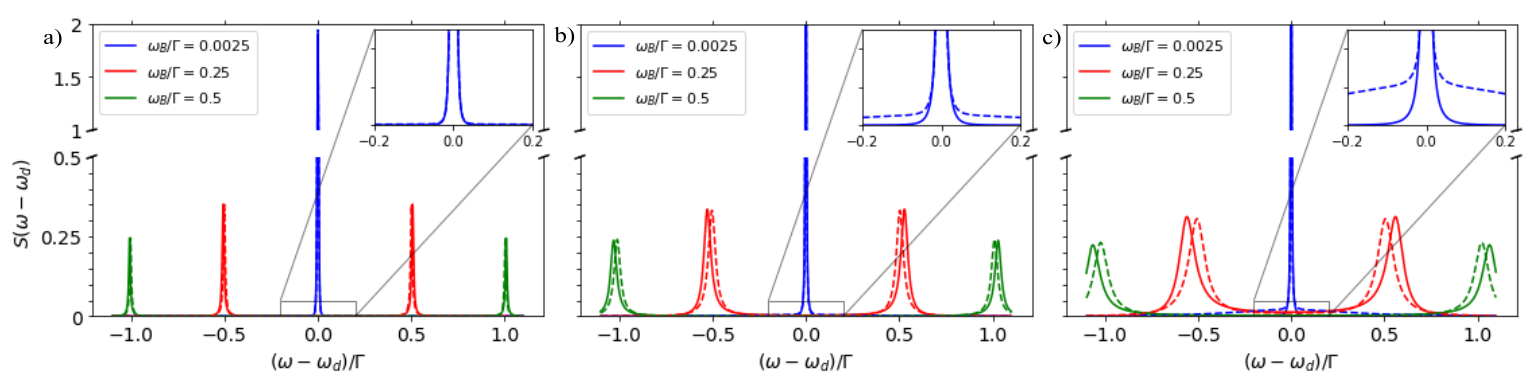}
\caption{Plots of analytic emission spectrum (Eq. \ref{eq: weak-driving spectrum}, solid lines) and numerically calculated emission spectrum, dashed lines, derived from solving Eq. \ref{eq: final Master equation} directly. \textbf{a)} Here $\Omega/\Gamma = 0.05$ and $\Delta/\Gamma = 0$. This plot shows that our analytic expression is in excellent agreement with the numerical solution for this parameter regime. This plot additionally highlights the distinct spectrum types with the spectrum appearing split for large magnetic fields but reduces to a single peak for small but non zero magnetic field strengths. \textbf{b)} Here $\Omega/\Gamma = 0.1$ and $\Delta/\Gamma = 0$. In this regime the weak excitation approximation begins to break down but the overall features of the spectrum are still captured. The discrepancy here is due to the lifetime broadened component of the scattered field that is neglected by the weak excitation approximation. \textbf{c)} Here $\Omega/\Gamma = 0.15$ and $\Delta/\Gamma = 0$. In this plot the lifetime broadened component is fairly significant as seen in the inset showing significant discrepancy between the two spectra.} 
\label{fig:Specs}
\end{figure*}

The analytical expression Eq. \ref{eq: weak-driving spectrum} is only valid in the weak excitation regime. For comparison to arbitrary driving powers we use a numerical solution to the full system master equation.  Examples are shown in Fig. \ref{fig:Specs}. Here, we have good agreement between the numerical and analytic methods when $\Omega$ is two orders of magnitude less than $\Gamma$ and the approximation breaks down as this difference approaches a single order of magnitude. This is when the lifetime-broadened component of the scattered field become significant and can no longer be neglected. This incoherent component can be 
recovered by including the matrix elements $\rho^{\prime}_{MP/PM}$ in Eq.~(\ref{eq: off-diagonal Lindblad equation}).

\end{document}